\documentclass[conference,compsoc]{IEEEtran}
\IEEEoverridecommandlockouts

\usepackage{tikz}
\usepackage{amsmath}
\usepackage{subcaption}
\usepackage{booktabs}
\usepackage{multirow}
\usepackage{cite}
\usepackage{dblfloatfix}
\usepackage{float}
\usepackage{subcaption}
\usepackage{relsize}
\usepackage{amsmath,mathtools}
\usepackage[hidelinks]{hyperref} 
\hypersetup{
    colorlinks=true, 
    linkcolor=blue, 
    urlcolor=blue, 
}
\usepackage{xurl}          
\usepackage{tabularx,booktabs,array}

\usepackage{listings}

\title{R+R: Revisiting Static Feature-Based Android Malware Detection using Machine Learning}

\author{\IEEEauthorblockN{1\textsuperscript{st} Md Tanvirul Alam}
\IEEEauthorblockA{\textit{Rochester Institute of Technology} \\
Rochester, NY, USA \\
ma8235@rit.edu}
\and
\IEEEauthorblockN{2\textsuperscript{nd} Dipkamal Bhusal}
\IEEEauthorblockA{\textit{Rochester Institute of Technology}\\
Rochester, NY, USA \\
db1702@rit.edu}
\and
\IEEEauthorblockN{3\textsuperscript{rd} Nidhi Rastogi}
\IEEEauthorblockA{\textit{Rochester Institute of Technology}\\
Rochester, NY, USA \\
nxrvse@rit.edu}

}

\def\month{MM}  
\def\year{YYYY} 

\begin{document}

\maketitle

\begin{abstract}
Static feature-based Android malware detection using machine learning (ML) remains critical due to its scalability and efficiency. However, existing approaches often overlook security-critical reproducibility concerns, such as dataset duplication, inadequate hyperparameter tuning, and variance from random initialization. This can significantly compromise the practical effectiveness of these systems. In this paper, we systematically investigate these challenges by proposing a more rigorous methodology for model selection and evaluation. Using two widely used datasets, Drebin and APIGraph, we evaluate six ML models of varying complexity under both offline and continuous active learning settings. Our analysis demonstrates that, contrary to popular belief, well-tuned, simpler models, particularly tree-based methods like XGBoost, consistently outperform more complex neural networks, especially when duplicates are removed. To promote transparency and reproducibility, we open-source our codebase, which is extensible for integrating new models and datasets, facilitating reproducible security research.
\end{abstract}

\section{Introduction}

The proliferation of Android malware continues to pose significant security threats, making timely detection and mitigation critical. Machine learning (ML), particularly leveraging static analysis features, remains widely adopted in security practice due to its scalability, speed, and suitability for rapidly screening large numbers of applications~\cite{arp2014drebin, mclaughlin2017deep, mamadroid,gopinath2023comprehensive}. ML models learn patterns from data, adapt to new threats, and scale far beyond the capabilities of traditional signature-based systems. However, recent research underscores critical limitations in their reliability, reproducibility, and security assurance~\cite {arp2022and}, leading to claims that cannot be reliably corroborated. Key issues, as observed in prior works~\cite{arp2022and, flood2024bad, verma2019data}, include sampling bias, incorrect labels, data snooping, inappropriate threat models, and unsuitable baselines. This lack of rigor in the design and evaluation process undermines the validity of many reported findings ~\cite{olszewski2023get}. Such discrepancies seriously question the reliability of ML-based malware detection when deployed in operational security settings.

\begin{figure}[t]
    \centering
    \includegraphics[width=0.35\textwidth]{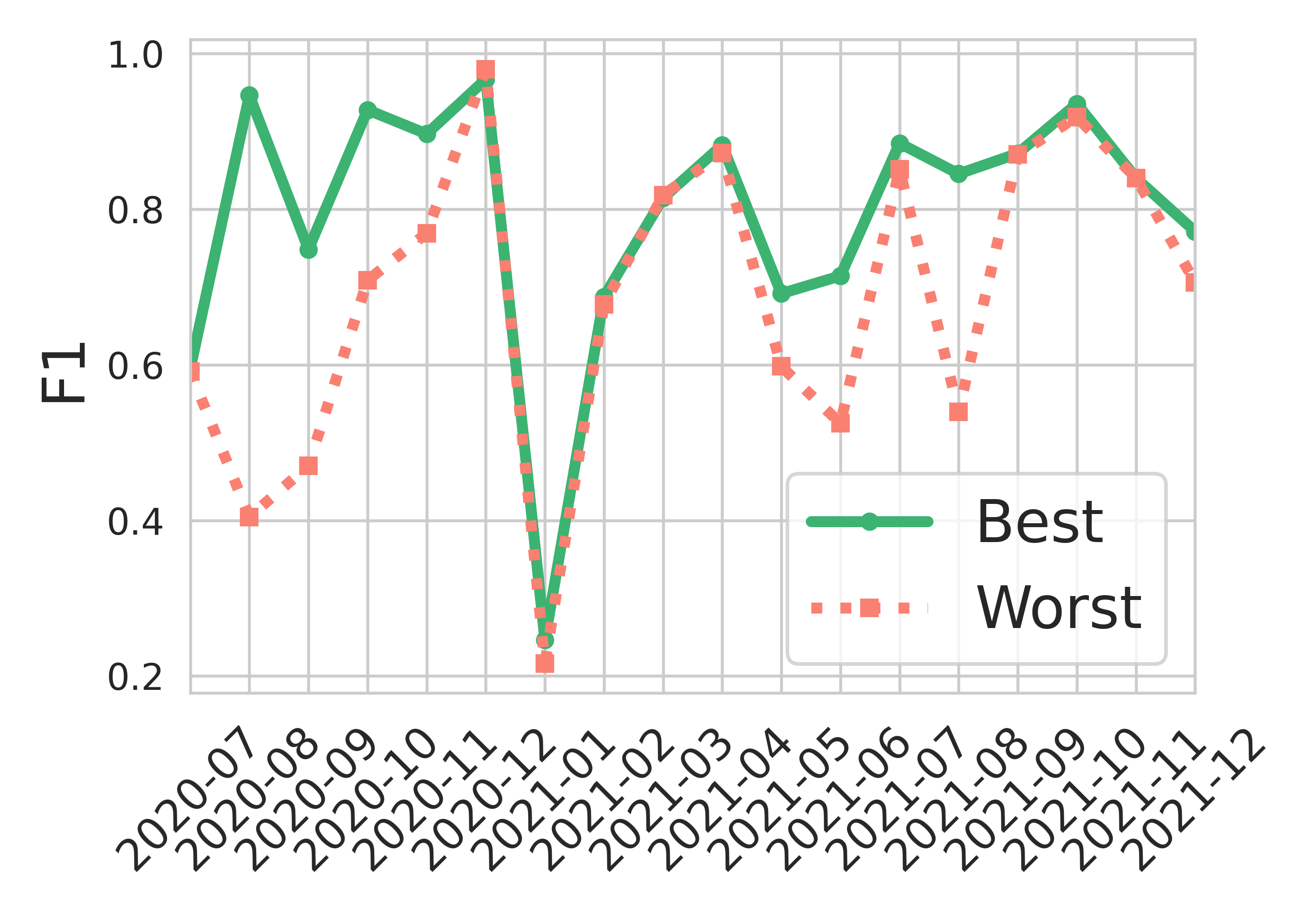}    \includegraphics[width=0.35\textwidth]{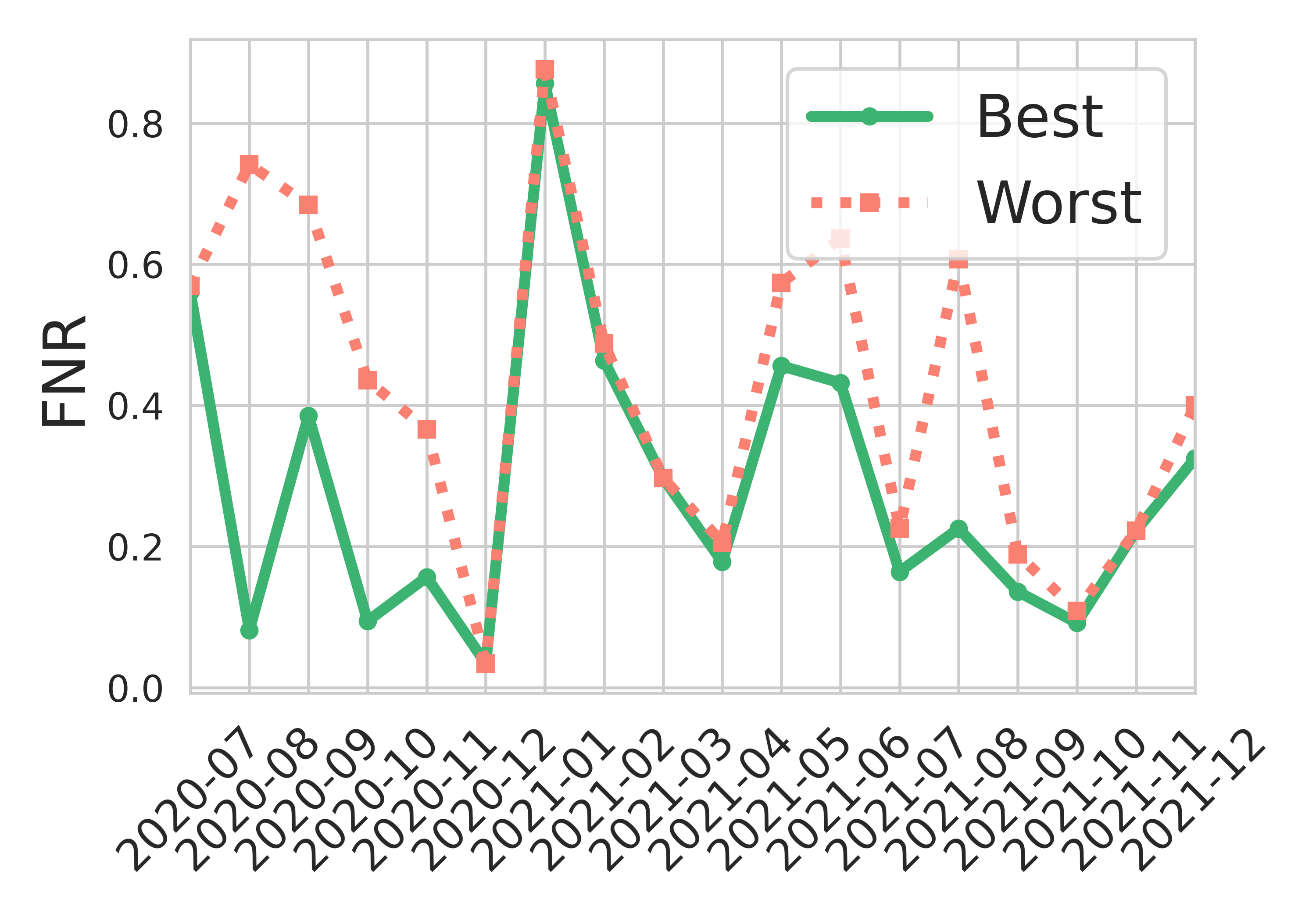}
    \caption{Performance comparison of the best and worst performing neural network models for Android malware detection using the state-of-the-art continuous active learning method from~\cite{chen2023continuous}. Top: F1-score. Bottom: False Negative Rate (FNR). Models were initialized with 5 different random seeds. The average F1-score over months differs by 10.6\%, and the False Negative Rate (FNR) differs by 13.9\% between the two models, despite using the same hyperparameters. This highlights reproducibility challenges in machine learning research for Android malware detection.}
    \label{fig:hcc-best-worst}
\end{figure}

\textbf{Motivation}. We showcase an example that highlights how reliably reproducing prior results in Android malware detection can be fraught with challenges, which, if not adequately addressed, can hinder progress in this research field. See Figure~\ref{fig:hcc-best-worst}, where we examine the reproducibility of results reported by~\cite{chen2023continuous} for Android malware detection using continuous active learning. Reproduced experiments include artifacts provided by the authors, including code and datasets, and the author-provided best set of hyperparameters for the Drebin-based dataset~\cite{arp2014drebin}. We used five different random seeds and then compared the best and worst-performing models on the test set. Best model achieved an average F1 score of 79.26\% over the test months, while the worst model scored 68.65\%, indicating a performance difference of 10.6\% solely due to variations in random seed initialization.

In this paper, we study these security-critical evaluation issues by systematically studying Android malware detection methodologies. We examine three prevalent reproducibility pitfalls, including (1) data duplication, which artificially inflates model performance; (2) inadequate hyperparameter tuning, causing unfair baseline comparisons; and (3) performance instability due to variance from random model initialization. We systematically discuss how these factors cause irreproducibility and propose ways to mitigate them. We use two widely-recognized Android malware datasets, Drebin~\cite{arp2014drebin} and APIGraph~\cite{zhang2020enhancing}, and six different models, including Random Forest~\cite{breiman2001random}, Support Vector Machine~\cite{arp2014drebin}, eXtreme Gradient Boosting~\cite{chen2016xgboost}, Multilayer Perceptron~\cite{rumelhart1986learning}, Supervised Contrastive Classifier~\cite{yang2021cade}, and Hierarchical Contrastive Classifier~\cite{chen2023continuous}, previously used in Android malware analysis. We conduct comprehensive experiments under realistic offline learning, where the model is trained once, and continuous active learning, where the model is periodically retrained with a subset of annotated samples. We perform detailed model selection and hyperparameter tuning for all models in both settings. While our work builds on recent advances in continuous active learning~\cite{chen2023continuous}, we broaden the evaluation scope to include diverse settings and emphasize reproducibility in Android malware detection. Contrary to the assumption that complex neural architectures outperform simpler ones, our experiments demonstrate that well-tuned tree-based models, particularly XGBoost, often exhibit greater robustness after duplicate samples are removed. These results underscore the importance of rigorous and reproducible evaluation over architectural sophistication.


\textbf{Key Contributions:}
\begin{enumerate}
    \item We systematically identify critical reproducibility issues in datasets and evaluation methods commonly used in Android malware detection, and propose rigorous methodological solutions to improve result reliability and enable fairer comparisons across models.
    \item Through comprehensive experiments on two standard datasets in both offline and continuous active learning scenarios, we demonstrate that simpler, well-tuned models (e.g., XGBoost) consistently outperform complex neural networks, challenging common assumptions in security literature.
    \item We release our extensively validated, extensible evaluation framework as open-source\footnote{\href{https://github.com/xashru/maldetect}{https://github.com/xashru/maldetect}}, facilitating transparent, reproducible, and security-focused malware detection research.

\end{enumerate}

\section{Background \& Related Work}

\begin{table*}[t]
\caption{Summary statistics of the datasets used in the study}
\centering

\begin{tabular}{@{}cccccccc@{}}
\toprule
\textbf{Dataset} & \textbf{Split} & \textbf{Duration} & \textbf{\begin{tabular}[c]{@{}c@{}}Benign \\ Apps\end{tabular}} & \textbf{\begin{tabular}[c]{@{}c@{}}Malicious\\ Apps\end{tabular}} & \textbf{Total} & \textbf{\begin{tabular}[c]{@{}c@{}}Malware\\ Families\end{tabular}} & \multicolumn{1}{l}{\textbf{\begin{tabular}[c]{@{}c@{}}Feature\\ Dimension\end{tabular}}} \\ \toprule
\multirow{3}{*}{Drebin} & Train & 2019-01 to 2019-12 & 40,947 & 4,542 & 45,489 & 121 & \multirow{3}{*}{16,978} \\
 & Validation & 2020-01 to 2020-06 & 18,109 & 2,028 & 20,137 & 67 &  \\
 & Test & 2020-07 to 2021-12 & 30,797 & 3,631 & 34,428 & 71 &  \\ \midrule
\multirow{3}{*}{APIGraph} & Train & 2012-01 to 2012-12 & 27,472 & 3,061 & 30,533 & 104 & \multirow{3}{*}{1,159} \\
 & Validation & 2013-01 to 2013-06 & 21,310 & 2,366 & 23,676 & 115 &  \\
 & Test & 2013-07 to 2018-12 & 240,729 & 25,377 & 266,106  & 418 &  \\ \bottomrule
\end{tabular}

\label{tab:datasets}
\end{table*}

\subsection{Reproducibility \& Replicability}

We adhere to ACM's definitions of reproducibility and replicability, standardized in 2020~\cite{acm:artifactReviewBadging}. Computational reproducibility refers to the ability of an independent team to achieve a study's results using the original study's artifacts~\cite{gundersen2018state}. This ensures that the reported findings are valid under specified conditions.~\cite{olszewski2023get} conducted a longitudinal study on the reproducibility of security papers published in Tier 1 venues (2013-2022), identifying challenges in reproducing results with author-provided artifacts and proposing mitigation strategies.

Replicability, in contrast, implies that an independent group can obtain the same results using artifacts they develop independently. Replicability studies generally involve different datasets and aim to confirm previous research results, considering the underlying system's inherent uncertainty.

Our study examines both aspects of machine learning research for Android malware detection. For instance, the analysis shown in Figure~\ref{fig:hcc-best-worst} underscores reproducibility issues in prior research, even when author-provided artifacts are used, due to unaccounted variability in published results. Additionally, we discuss pitfalls that must be addressed for a study's claims to be replicable with independently developed artifacts. In this paper, we use the term reproducibility to broadly refer to both computational reproducibility and replicability.



\subsection{Android Malware Detection}

Android malware detection using machine learning classifies applications as benign or malicious~\cite{liu2020review}. Three main types of features are used: static, dynamic, and hybrid. Static features are obtained by analyzing the app's source code or related information~\cite{arp2014drebin,mamadroid,zhang2020enhancing}. Specifically, the primary focus for Android applications is the APK file, the installation package, which includes components like AndroidManifest.xml and smali files obtained through decompilation. Dynamic features are acquired by observing the app's behavior in real or emulated environments, such as sandboxes~\cite{jannat2019analysis,shankar2017androtaint,zhang2018dalvik}. Hybrid features combine both static and dynamic characteristics~\cite{choudhary2018haamd,martinelli2017bridemaid}. 

Extracted features are input into machine learning models like Random Forests~\cite{breiman2001random}, Support Vector Machines~\cite{arp2014drebin}, and Multilayer Perceptrons~\cite{rumelhart1986learning} to classify apps as malware or benign. The choice of features and models depends on detection objectives and resources. Static analysis is fast and suitable for large-scale detection, whereas dynamic analysis can be more accurate but resource-intensive~\cite{aghakhani2020malware}. This work focuses on static-feature-based machine learning methods due to their prevalence in the literature~\cite{ jordaney2017transcend, pendlebury2019tesseract, zhang2020enhancing, barbero2022transcending, chen2023continuous}.

\textbf{Concept Drift:} Malware detection systems encounter challenges due to the evolving nature of malware, leading to concept drift. This drift can arise from new malware families, behavioral changes, evasion attempts, or updates in API semantics. Traditional supervised learning models trained on static datasets may become outdated as new malware types emerge, reducing detection accuracy~\cite{yang2021cade,chow2023drift,pendlebury2019tesseract,xu2019droidevolver}. These models often fail to recognize emerging variants, resulting in higher false negatives. To address this, continuous learning methods can enable models to adapt to new data continuously and stay current with malware trends~\cite{jordaney2017transcend,chen2023continuous}.

\textbf{Reproducibility of Research in Android Malware Detection Using ML:} Previous studies have assessed issues in reported results for Android malware detection~\cite{zhao2021impact,irolla2018duplication}, attempted to replicate prior research~\cite{daoudi2021lessons}, or addressed limitations in practical environments~\cite{gao2024comprehensive}. However, these studies do not consider the continuous learning setup, often rely on single global metrics like AUC or F1-score, lacking nuanced analysis of monthly performance variations, factors influencing reproducibility, and inherent variances in machine learning models due to their stochastic nature. Our work aims to address these gaps by analyzing conditions that can lead to poor reproducibility in different experimental settings and how to mitigate them.

\section{Study settings}


\subsection{Machine Learning Scenarios}
We evaluate Android malware detection models under 2 distinct ML scenarios relevant to practical security deployment:

\begin{enumerate}
    \item \textbf{Offline Learning:} Models are trained once on the historical dataset and then evaluated on unseen future data, reflecting static deployment scenarios. This involves using annotated data from a specific period (e.g., one year) to train a model.

    \item \textbf{Continuous Active Learning:} To address the evolving nature of malware (concept drift), we periodically retrain the model~\cite{jordaney2017transcend,zhang2020enhancing,chen2023continuous} using only the samples where the model is least confident, saving annotation resources. We adopt monthly retraining intervals, consistent with established literature.
\end{enumerate}

We treat these settings separately because each requires different model selection methods, such as hyperparameter tuning, as noted in prior work~\cite{chen2023continuous}. 

\subsection{Datasets}
We utilize the publicly available dataset from~\cite{chen2023continuous}, which includes two datasets based on static feature types: Drebin~\cite{arp2014drebin} and APIGraph~\cite{zhang2020enhancing}. For brevity, we refer to them as the Drebin and APIGraph datasets, although the Drebin variant used here was collected between 2019 and 2021, rather than the original 2010–2012 corpus~\cite{arp2014drebin}. This ensures comparability with the evaluation setup of~\cite{chen2023continuous}. Both datasets have been extensively applied across various contexts, including explainable ML~\cite{liu2022explainable}, concept drift analysis~\cite{gao2024comprehensive,barbero2022transcending}, obfuscation studies~\cite{gao2023obfuscation}, and semi-supervised learning~\cite{kan2021investigating,alam2024morph}, reflecting their central role in the literature. However, despite their broad use, prior studies have not adequately examined the inherent reproducibility challenges in these datasets, which complicates efforts to replicate results in new environments. Addressing these issues provides the primary motivation for our work.

Drebin uses eight different sets of features, including access to hardware components, requested permissions, app components, intents, usage of restricted API calls, used permissions, suspicious API calls, and network addresses. APIGraph employs API semantics to cluster similar APIs, significantly reducing the feature space. The resulting feature set is binary for both datasets, indicating whether a feature is present or absent for a specific sample.

The Drebin dataset contains applications collected from 2019 to 2021, while the APIGraph dataset includes applications from 2012 to 2018. Both datasets address temporal and spatial biases present in malware datasets~\cite{arp2022and, pendlebury2019tesseract}. They consist of approximately 90\% benign and 10\% malware applications collected from each month, reflecting their real-world ratios. Samples are ordered and approximately evenly distributed over their respective periods. Table~\ref{tab:datasets} provides a summary of the datasets.

Following~\cite{chen2023continuous}, we use data from the first year as the training set, the next six months for validation, and the remaining months for testing. For offline learning, only the training set data is used to train the model. In the continuous active learning setup, this data is used to train the initial model, which is then periodically retrained using data from subsequent months from the validation and test set. The validation set is used to tune the model hyperparameters in both experimental settings.

\begin{figure*}[t] 
\centering
\begin{subfigure}{0.47\textwidth} 
    \includegraphics[width=\textwidth]{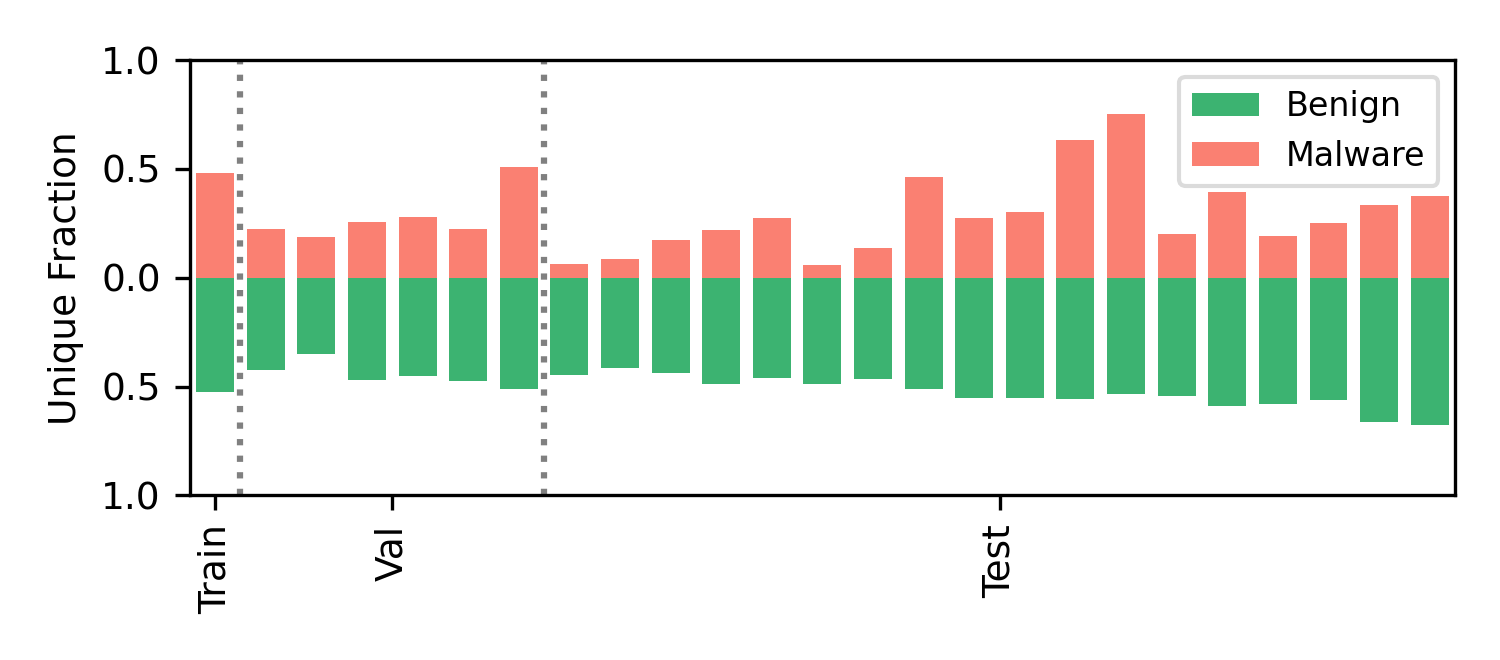} 
    \caption{Drebin: Offline learning deduplication}
    \label{fig:image1}
\end{subfigure}
\begin{subfigure}{0.47\textwidth}
    \includegraphics[width=\textwidth]{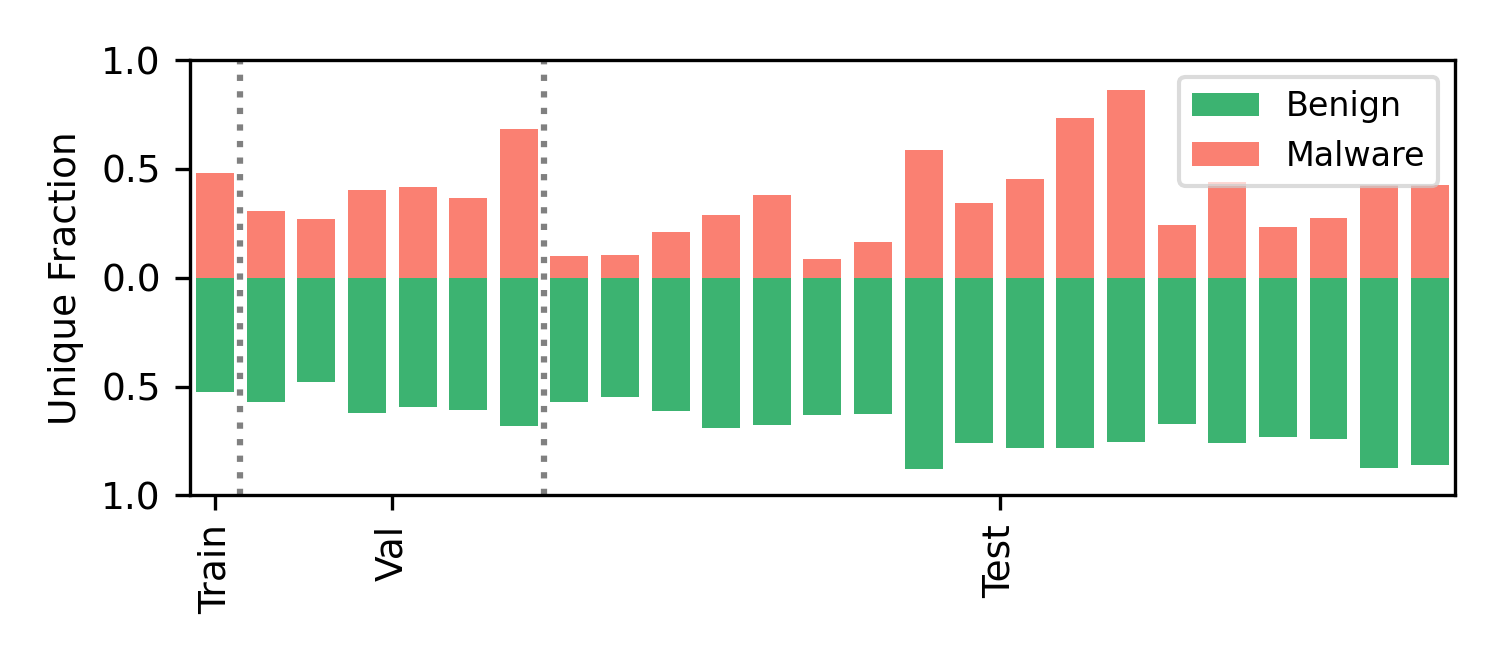}
    \caption{Drebin: Continuous active learning deduplication}
    \label{fig:image2}
\end{subfigure}

\vspace{-0pt}  

\begin{subfigure}{0.47\textwidth}
    \includegraphics[width=\textwidth]{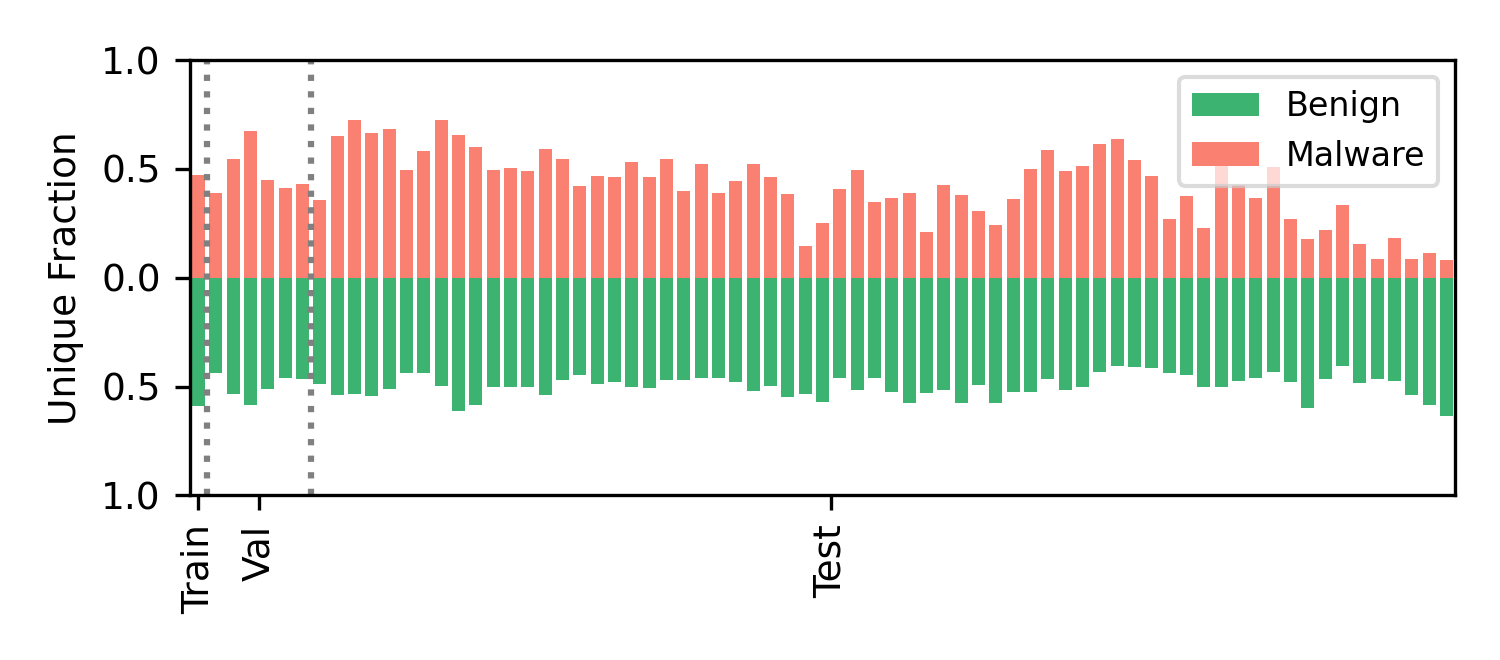}
    \caption{APIGraph: Offline learning deduplication}
    \label{fig:image3}
\end{subfigure}
\begin{subfigure}{0.47\textwidth}
    \includegraphics[width=\textwidth]{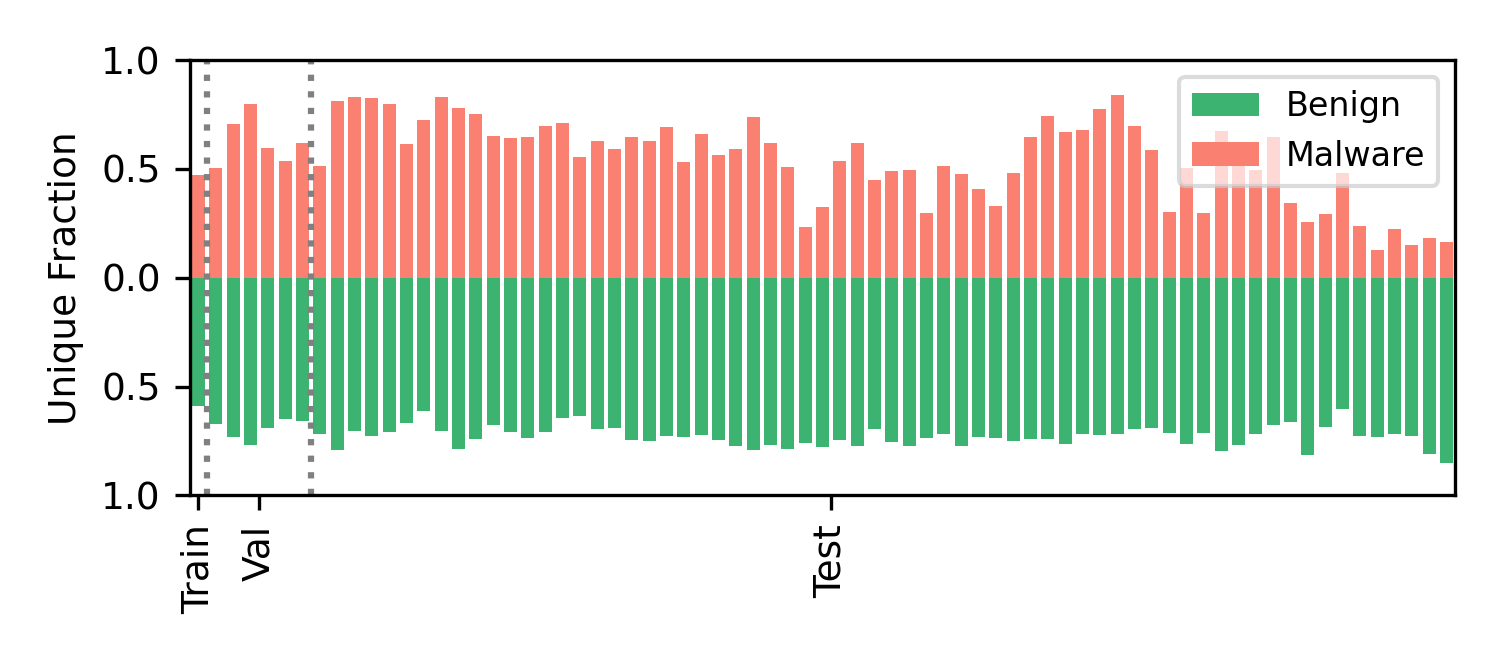}
    \caption{APIGraph: Continuous active learning deduplication}
    \label{fig:image4}
\end{subfigure}

\caption{Fraction of unique samples retained after the deduplication process}
\label{fig:duplicates}
\end{figure*}

\section{Pitfalls in Reproducibility and Evaluation}\label{sec:pitfalls}
We critically examine key methodological and dataset-related pitfalls prevalent in existing Android malware detection research. These pitfalls can severely undermine reproducibility and practical security effectiveness, and we propose systematic strategies to mitigate their impact.

\subsection{Dataset Duplication} 
Issues of duplicates in Android malware datasets have been reported in prior studies~\cite{irolla2018duplication,zhao2021impact}. The study by~\cite{irolla2018duplication} analyzed duplicates in the original Drebin dataset~\cite{arp2014drebin} using opcode sequences and found that 50.65\% of malware samples were unique, with the rest being duplicates. They demonstrated that removing duplicates can alter the performance rankings of different machine learning models. \cite{zhao2021impact} examined duplicates in four malware datasets using APK, DEX, opcode sequences, and API calls to identify them. They also evaluated the impact of duplicates on multiple machine-learning models.

Our study differs from prior works in key ways. First, we are the first to address the reproducibility of reported results in this context. Second, we provide a time-aware analysis of duplicates' effects on malware detection models, as global metrics (e.g., F1-score) may miss monthly variations~\cite{pendlebury2019tesseract}. Lastly, unlike prior studies focused only on offline learning, we assess duplicates' impact in offline and continuous learning settings, which require different deduplication approaches, as discussed later.

\subsubsection{Defining Duplicates}

We define duplicates based strictly on feature-vector equality, meaning two samples are duplicates if their extracted feature vectors are identical. Specifically, two samples \( X \) and \( Z \) are considered duplicates if \( x_1 = z_1, x_2 = z_2, \ldots, x_n = z_n \) for an \( n \)-dimensional feature vector:
\[
X = (x_1, x_2, \ldots, x_n), \quad Z = (z_1, z_2, \ldots, z_n)
\]
This definition directly reflects the model’s perspective, differing from prior approaches relying on raw or opcode-based duplication checks~\cite{irolla2018duplication,zhao2021impact}. 
Notably, despite using this feature-space definition, we observe a similar percentage of duplicates in the datasets as reported in prior studies. Also, detecting accurate duplicates is particularly challenging in malware analysis, which we further discuss in Section~\ref{sec:limitations}.

\subsubsection{Deduplication} \label{sec-dedup}

The process of \textit{deduplication} ensures that the dataset contains unique samples. We establish rigorous deduplication protocols based on different experimental settings in our study:

\textbf{Offline Learning Deduplication:} Here, the model is trained once on a training set and evaluated on validation and test sets. Deduplication involves selecting unique samples within and across data splits. Since the model is evaluated monthly, two levels of deduplication are required. \textit{First}, we remove samples that appear in earlier data splits: duplicates in the validation set are removed if present in the training set, while the test set excludes duplicates from both the training and validation sets. \textit{Second}, within each split, only the earliest occurrence of a sample is retained based on its appearance time. 

\textbf{Continuous Active Learning Deduplication:} Here, the definition of duplicates evolves due to monthly model retraining and changes in behavior, such as catastrophic forgetting~\cite{rahman2022limitations}. Duplicate samples are first removed from the training split, following the same approach as offline learning. For the validation and test sets, duplicates are removed only within the same month, not across different months. This allows duplicates to appear in multiple months of the test set while ensuring that each month contains unique instances. This approach is crucial for continuous learning, as periodic retraining requires the model to recognize recurring samples accurately.

\subsubsection{Duplicate Statistics in the Datasets}

Figure~\ref{fig:duplicates} illustrates the proportion of unique samples in the datasets after the deduplication process in both experimental settings. Ideally, this fraction would be 1 for both malware and benign apps, indicating no duplicates. However, our analysis shows that many duplicates are present, with varying degrees across different splits. Notably, around 50\% of the training set for the Drebin and APIGraph datasets consists of duplicate samples in benign and malware categories.

We observe two key patterns regarding duplicates in the validation and test sets. First, the fraction of duplicate samples varies monthly. Second, malware samples tend to contain more duplicates than benign samples, with certain months particularly affected. For instance, in the first test month of the Drebin dataset (July 2020), only 6.30\% of malware samples are unique in the offline-learning setting, and 9.96\% are unique in the continuous-learning setting. 


\subsection{Inadequate Model Selection and Tuning}

A common approach for evaluating machine learning models involves splitting the dataset into training, validation, and test sets. The validation set is essential for model selection and hyperparameter tuning to prevent biased parameter choices~\cite{arp2022and}. However, previous research on Android malware detection often omits a separate validation set, leading to incomplete evaluations~\cite{arp2022and}. Our study adopts the same train, validation, and test split as in~\cite{chen2023continuous} for offline and continuous learning settings.

Another common issue is the inadequate tuning of baseline methods~\cite{arp2022and}. This often results in insufficient hyperparameter tuning for simpler models compared to more complex ones. For example, prior research has shown that appropriately tuned tree-based methods frequently outperform deep neural networks on tabular datasets~\cite{grinsztajn2022tree,mcelfresh2024neural}, even though neural networks excel in tasks like computer vision and natural language processing~\cite{he2016deep,DBLP:conf/nips/VaswaniSPUJGKP17,DBLP:conf/naacl/DevlinCLT19}.

In Android malware detection, various models have been used in literature, such as support vector machines (SVM), random forests, gradient-boosted decision trees, and deep neural networks~\cite{arp2014drebin,mamadroid,pendlebury2019tesseract,zhang2020enhancing,chen2023continuous}. However, these models might not have been adequately calibrated due to the absence of a validation set or insufficient hyperparameter searches. Our analysis emphasizes extensive hyperparameter searches across different baseline methods, ensuring the same hyperparameter budget is allocated to each model. This approach enables a fair comparison between models and aligns with the established norms in the domain generalization literature in machine learning~\cite{domainbed}.


\subsection{Accounting for Variance from Random Initialization}

Some machine learning models, particularly neural networks, exhibit additional stochasticity independent of their hyperparameters. Their sensitivity to initial random weight initialization can lead to variations in results that depend on the training system rather than the model parameters~\cite{bouthillier2021accounting,picard2021torch}. It is common practice in the deep learning community to evaluate performance across multiple random seeds on the test set and report the average~\cite{bouthillier2020survey} to address this. However, prior studies on Android malware detection often neglect this variation, leading to irreproducible outcomes when experiments are conducted with different random seeds. We explore this issue in detail in Section~\ref{sec-offline-variance}.

\subsection{Delayed Evaluation of Models} \label{sec-delayed}
Splitting the dataset into non-overlapping temporal train, validation, and test sets enables consistent model selection but introduces a delay between training and test data. While acceptable in standard supervised learning, this temporal gap poses challenges for malware detection, where data distributions evolve. As observed in our study and in~\cite{chen2023continuous}, a six-month delay between training and test periods can lead to unrealistic evaluations, since deployed models would typically be retrained with more recent data~\cite{arp2022and}.

To mitigate this, we merge the training and validation datasets after hyperparameter tuning and evaluate on the test set. This \textbf{Merged Training} setup offers a fairer estimate of offline performance compared to the \textbf{Holdout Training} setup, where only the training data are used. Such merging is standard in temporal domain generalization~\cite{yao2022wild}, and the impact of both setups is analyzed in Sections~\ref{sec-offline-res} and~\ref{sec-active-res}.

\section{Experiments}

\subsection{Machine Learning Models}

We utilize six different models commonly used in prior works on Android malware detection:

\begin{enumerate}
\item \textbf{Random Forest (RF):} Random Forest~\cite{breiman2001random} is an ensemble learning method for classification that constructs multiple decision trees during training and outputs the mode of their classes. It has been used in Android malware detection~\cite{sanz2013puma,mamadroid}.

\item \textbf{Support Vector Machine (SVM):} SVM~\cite{cortes1995support} is a classification algorithm that finds the hyperplane that best separates classes in the feature space. It has been widely used for Android malware detection, particularly with Drebin features~\cite{arp2014drebin,jordaney2017transcend}.

\item \textbf{eXtreme Gradient Boosting (XGBoost):} XGBoost~\cite{chen2016xgboost} is a scalable and efficient gradient boosting method that excels in performance on various tabular datasets~\cite{grinsztajn2022tree}, although it has not shown consistent success in Android malware detection~\cite{chen2023continuous}.

\item \textbf{Multilayer Perceptron (MLP):} An MLP is a type of neural network composed of multiple layers of neurons, with each layer fully connected to the next~\cite{rumelhart1986learning}. MLPs are used for a variety of tasks, including classification and regression, and have been employed in several prior works for Android malware detection~\cite{pendlebury2019tesseract,yuan2016droiddetector,kim2018multimodal}.

\item \textbf{Supervised Contrastive Classifier (SCC):} Contrastive learning has proven effective for detecting and adapting to concept drift in Android malware detection~\cite{yang2021cade,chen2023continuous}. It learns representations by bringing similar instances closer and pushing dissimilar ones apart~\cite{wu2018unsupervised,khosla2020supervised}. Our implementation uses a triplet loss over the two main classes, benign and malware, where positives share the same label as the anchor and negatives differ. Unlike prior works that model each malware family as a separate class, our binary setup is more annotation efficient and better suited for real world detection.

The learning process combines a supervised binary cross-entropy loss and a triplet margin loss. The binary cross-entropy loss is defined as:

\[
\mathcal{L}_{\text{bce}} = -\frac{1}{N} \sum_{i=1}^{N} \left[ y_i \log(\hat{y}_i) + (1-y_i) \log(1-\hat{y}_i) \right]
\]

where \( y_i \) is the true label and \( \hat{y}_i \) is the predicted probability.

The triplet margin loss is given by:


\[
\begin{aligned}
\mathcal{L}_{\text{triplet}}
&= \frac{1}{N} \sum_{i=1}^{N}
   \max\Bigl(0,\;
   \|\mathrm{enc}(x_i^{a}) - \mathrm{enc}(x_i^{p})\|_2^{2} \\
&\quad \mathllap{-}\; \|\mathrm{enc}(x_i^{a}) - \mathrm{enc}(x_i^{n})\|_2^{2} + m \Bigr)
\end{aligned}
\]

where \(x_i^a, x_i^p, x_i^n\) are the anchor, positive, and negative samples respectively, \(\text{enc}(\cdot)\) is the encoder function, and \(m\) is the margin parameter.

The combined loss function is:

\[
\mathcal{L} = \lambda \cdot \mathcal{L}_{\text{bce}} +  \mathcal{L}_{\text{triplet}}
\]

where \(\lambda\) is a weight parameter balancing the two loss components.

\item \textbf{Hierarchical Contrastive Classifier (HCC):} Following~\cite{chen2023continuous}, HCC extends supervised contrastive learning with a hierarchical loss that enforces stronger similarity among samples from the same malware family than those from different families. Using an encoder \( \mathrm{enc} \), embeddings of benign or cross-family malware pairs satisfy \( \|\mathrm{enc}(x_1) - \mathrm{enc}(x_2)\|_2 \leq m \), while benign–malicious pairs are pushed farther apart with \( \|\mathrm{enc}(x_1) - \mathrm{enc}(x_2)\|_2 \geq 2m \).

Let \( d_{ij} \) denote the Euclidean distance between two samples \( i \) and \( j \) in the embedding space, defined as \( d_{ij} = \|\text{enc}(x_i) - \text{enc}(x_j)\|_2 \). The hierarchical contrastive loss is:

\[
\begin{aligned}
\mathcal{L}_{hc}(i) = & \frac{1}{|P(i, y_i, y_i')|} \sum_{j \in P(i, y_i, y_i')} \max(0, d_{ij} - m) \\
& + \frac{1}{|P_z(i, y_i, y_i')|} \sum_{j \in P_z(i, y_i, y_i')} d_{ij} \\
& + \frac{1}{|N(i, y_i)|} \sum_{j \in N(i, y_i)} \max(0, 2m - d_{ij})
\end{aligned}
\]

The first term ensures positive pairs, such as (benign, benign) or (malicious, malicious), are close but not overly constrained, penalizing distances only if they exceed \( m \). The second term enforces similarity within the same malware family. The last term aims to separate benign and malicious samples by at least \( 2m \).

Like the SCC, this loss is combined with binary cross-entropy loss to train the model. Note that this is the only model that requires access to malware family labels to train the model.

\end{enumerate}

\subsection{Active Learning Sample Selection Strategy} \label{sec-al-sample}



Active learning seeks to improve model performance with minimal labeled data by selectively querying the most informative samples from a large pool of unlabeled data~\cite{ren2021survey}. For all models except HCC, we employ the standard uncertainty sampling strategy~\cite{DBLP:conf/sigir/LewisG94}, which selects samples where the model is least confident, computed as \(1 - \text{probability}\) (e.g., the softmax output for binary classifiers). For HCC, we follow the pseudo-label selection method of~\cite{chen2023continuous}, where pseudo-labels act as ground truth for calculating a contrastive pseudo-loss between test and nearby training embeddings. This loss is combined with binary cross-entropy to identify samples with the highest uncertainty.

\begin{table*}[ht]
\centering
\caption{Performance comparison of different models on the Drebin dataset for offline learning. The results represent the mean and standard deviation over five runs with different random seeds. The best-performing model is underlined.}

\resizebox{\textwidth}{!}{%
\setlength{\tabcolsep}{3.25pt}
\begin{tabular}{l|ccc|ccc|ccc|ccc} 
\toprule
\multirow{2}{*}{\textbf{Model}} & \multicolumn{6}{c|}{\textbf{Merged Training}} & \multicolumn{6}{c}{\textbf{Holdout Training}} \\
\cmidrule(lr){2-7} \cmidrule(lr){8-13}
& \multicolumn{3}{c|}{\textbf{Duplicated}} & \multicolumn{3}{c|}{\textbf{Deduplicated}} & \multicolumn{3}{c|}{\textbf{Duplicated}} & \multicolumn{3}{c}{\textbf{Deduplicated}} \\
\cmidrule(lr){2-4} \cmidrule(lr){5-7} \cmidrule(lr){8-10} \cmidrule(lr){11-13}
& F1 & FPR & FNR & F1 & FPR & FNR & F1 & FPR & FNR & F1 & FPR & FNR \\
\midrule
RF & 46.7±0.99 & \underline{0.14±0.00} & 66.8±0.74 & 41.5±0.30 & \underline{0.08±0.00} & 72.5±0.25 & 39.4±1.31 & \underline{0.16±0.03} & 72.4±1.03 & 33.2±1.01 & \underline{0.14±0.04} & 78.7±0.71 \\ 
SVM & 63.6±0.00 & 0.80±0.00 & 47.0±0.00 & 46.6±0.00 & 0.60±0.00 & 65.6±0.00 & 47.2±0.00 & 0.92±0.00 & \underline{63.2±0.00} & 33.8±0.00 & 0.70±0.00 & 76.7±0.00 \\ 
XGBoost & 63.7±2.45 & 0.18±0.01 & 48.9±2.80 & \underline{59.5±0.46} & 0.39±0.02 & 52.1±0.59 & \underline{45.2±1.92} & 0.24±0.04 & 67.3±2.01 & \underline{47.3±0.78} & 0.73±0.02 & \underline{63.4±0.66} \\ 
MLP & 66.2±3.69 & 0.41±0.08 & 45.8±4.10 & 53.6±2.08 & 0.39±0.08 & 59.5±2.12 & 44.3±2.07 & 0.87±0.06 & 66.7±1.72 & 40.9±1.05 & 0.57±0.04 & 69.8±0.82 \\ 
SCC & \underline{74.7±0.78} & 0.53±0.09 & \underline{35.0±1.51} & 57.4±0.44 & 0.63±0.25 & 53.7±2.11 & 45.6±1.18 & 0.70±0.11 & 65.7±1.35 & 37.3±0.26 & 0.63±0.10 & 73.0±0.61 \\ 
HCC & 68.4±1.73 & 0.82±0.21 & 41.3±2.52 & 56.7±0.79 & 1.04±0.12 & \underline{50.5±0.76} & 44.9±1.60 & 0.77±0.11 & 66.2±1.60 & 35.8±0.65 & 0.84±0.15 & 73.5±1.14 \\ 
\bottomrule
\end{tabular}
}

\label{tab:res-static-drebin}
\end{table*}

\begin{table*}[ht]
\centering
\caption{Performance comparison of different models on the APIGraph Dataset for offline learning}

\resizebox{\textwidth}{!}{%
\setlength{\tabcolsep}{3.25pt}
\begin{tabular}{l|ccc|ccc|ccc|ccc} 
\toprule

\multirow{2}{*}{\textbf{Model}} & \multicolumn{6}{c|}{\textbf{Merged Training}} & \multicolumn{6}{c}{\textbf{Holdout Training}} \\
\cmidrule(lr){2-7} \cmidrule(lr){8-13}
& \multicolumn{3}{c|}{\textbf{Duplicated}} & \multicolumn{3}{c|}{\textbf{Deduplicated}} & \multicolumn{3}{c|}{\textbf{Duplicated}} & \multicolumn{3}{c}{\textbf{Deduplicated}} \\
\cmidrule(lr){2-4} \cmidrule(lr){5-7} \cmidrule(lr){8-10} \cmidrule(lr){11-13}
& F1 & FPR & FNR & F1 & FPR & FNR & F1 & FPR & FNR & F1 & FPR & FNR \\
\midrule
RF & 64.6±1.02 & \underline{0.23±0.00} & 49.6±1.08 & 61.3±0.92 & \underline{0.50±0.03} & 51.7±0.79 & 45.3±5.22 & \underline{0.12±0.02} & 68.5±4.50 & 46.8±3.89 & \underline{0.34±0.08} & 67.0±3.76  \\ 
SVM & \underline{76.0±0.00} & 1.20±0.00 & \underline{30.7±0.00} & \underline{69.7±0.00} & 1.41±0.00 & 34.9±0.00 & \underline{74.5±0.00} & 0.86±0.00 & \underline{34.7±0.00} & \underline{66.3±0.00} & 1.42±0.00 & \underline{39.5±0.00}  \\ 
XGBoost & 74.6±0.47 & 1.21±0.09 & 32.5±0.49 & 69.0±0.45 & 1.61±0.03 & \underline{34.3±0.57} & 69.5±1.40 & 1.03±0.06 & 40.2±1.65 & 65.2±0.62 & 1.33±0.05 & 41.6±0.88  \\ 
MLP & 68.6±2.28 & 0.85±0.26 & 41.9±3.77 & 63.0±2.73 & 0.96±0.37 & 46.32±4.9 & 55.6±4.26 & 0.90±0.20 & 56.2±4.85 & 57.0±3.52 & 1.67±0.64 & 49.9±6.78  \\ 
SCC & 66.1±2.79 & 1.06±0.26 & 43.8±3.94 & 66.2±2.09 & 2.11±0.57 & 35.2±1.98 & 70.36±2.9 & 1.79±0.16 & 35.3±3.33 & 57.0±3.52 & 1.67±0.64 & 49.9±6.78  \\ 
HCC & 73.3±2.37 & 1.35±0.15 & 33.3±3.90 & 66.0±1.75 & 0.92±0.17 & 42.8±2.66 & 64.2±5.17 & 1.60±0.20 & 43.8±6.61 & 62.6±2.74 & 1.59±0.26 & 43.8±4.13 \\ 
\bottomrule
\end{tabular}
}

\label{tab:res-static-apigraph}
\end{table*}

\subsection{Hyperparameter Tuning}\label{sec-hopt}





We perform hyperparameter tuning independently for two experimental setups: offline learning and continuous active learning. In the offline setup, each model is trained once on the training dataset, using the validation set to select optimal hyperparameters. We apply random search~\cite{bergstra2012random} with a budget of 200 trials per model, ensuring fair comparison across models and accounting for differences in search spaces, and repeat tuning separately for duplicated and deduplicated datasets. In the continuous active learning setup, following~\cite{chen2023continuous}, the validation data is divided into six-month intervals; the model is retrained with active learning at each interval and evaluated on the subsequent month. Most hyperparameters remain fixed across retraining phases, though parameters such as training epochs may vary. Random search with 100 trials per model and an annotation budget of 50 samples per month is used for tuning. Full hyperparameter details are provided in Appendix~\ref{appendix-hpo}.

\subsection{Evaluation Metric}

After hyperparameter tuning, we evaluate the models on the test set. The continuous active learning setup includes retraining the model each test month and evaluating it the following month. We use the same performance metrics as in~\cite{chen2023continuous}: average F1-score, False Positive Rate (FPR), and False Negative Rate (FNR). We evaluate all models on the test set using optimal hyperparameters and five different random seeds, reporting the mean and standard deviation of the metrics. It is worth noting that SVM produces deterministic results across seeds, as it uses the same training dataset without any stochastic elements.




\subsection{Experimental Settings}
We summarize the different experimental settings below:

\begin{enumerate}
\item \textbf{Learning Settings:}
\begin{itemize}
    \item \textit{Offline learning}: The model is trained once.
    \item \textit{Active learning}: The model is retrained monthly with a fixed annotation budget. For RF, SVM, and XGBoost, we train from scratch, while for neural networks, we continue fine-tuning the previous model, following recommendations from~\cite{chen2023continuous}.
\end{itemize}

    \item \textbf{Duplicates:} 
    \begin{itemize}
        \item \textit{Duplicated}: We use the original datasets containing duplicated samples in the train, validation, and test splits.
        \item \textit{Deduplicated}: We deduplicate the datasets following the procedures in Section~\ref{sec-dedup} before conducting the experiments.
    \end{itemize}

    \item \textbf{Delayed Evaluation:} Based on whether validation data is merged with training data after hyperparameter tuning(\ref{sec-delayed}):
    \begin{itemize}
        \item \textit{Merged Training}: Merge validation data with training data before evaluating the model on test data.
        \item \textit{Holdout Training}: Train the model using only training data.
    \end{itemize}

\item \textbf{Active Learning Budget:} We evaluate two monthly annotation budgets for the active learning setup: 50 and 100 samples. Both settings use the model optimized with a 50-sample budget for hyperparameter tuning.

\end{enumerate}

\begin{figure*}
\centering
    \includegraphics[width=0.49\textwidth]{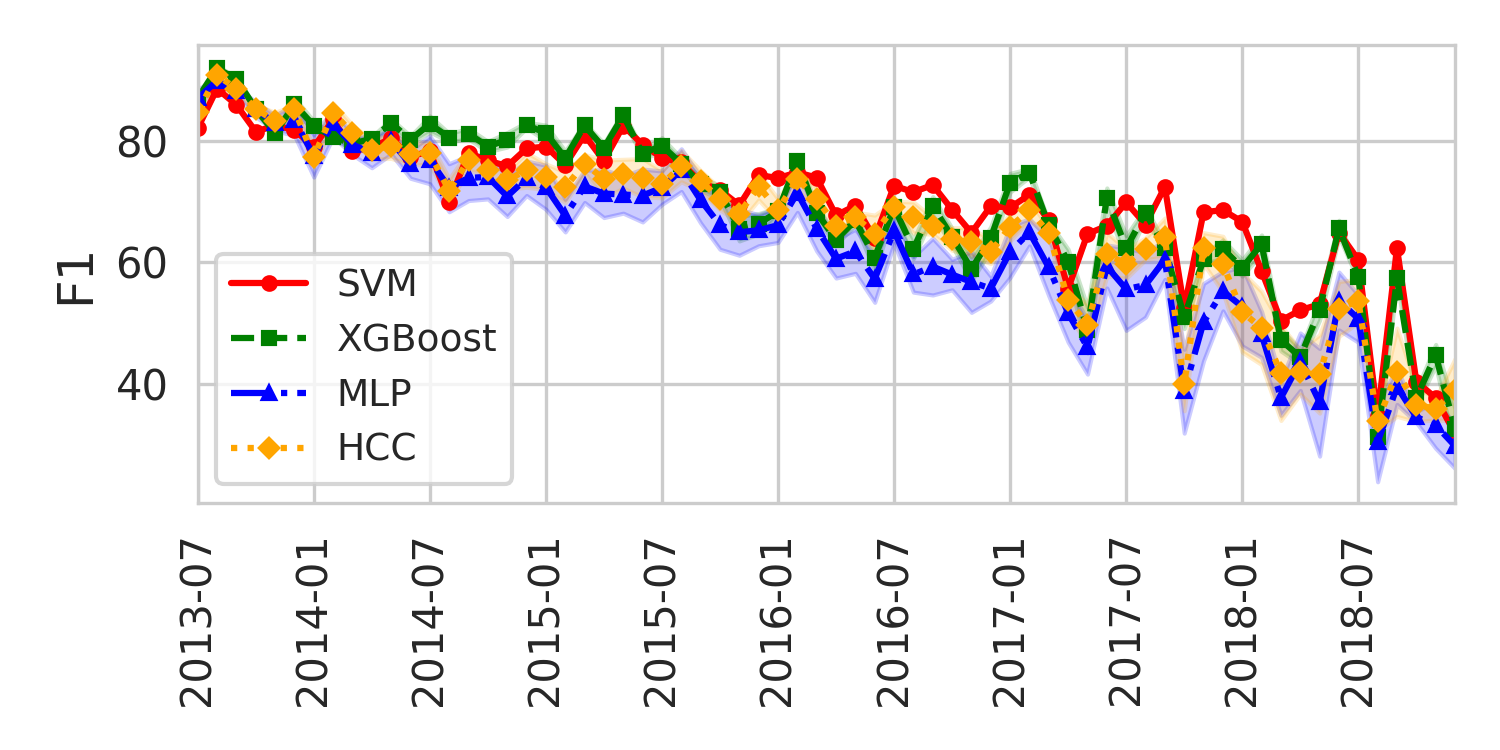}
    \includegraphics[width=0.49\textwidth]{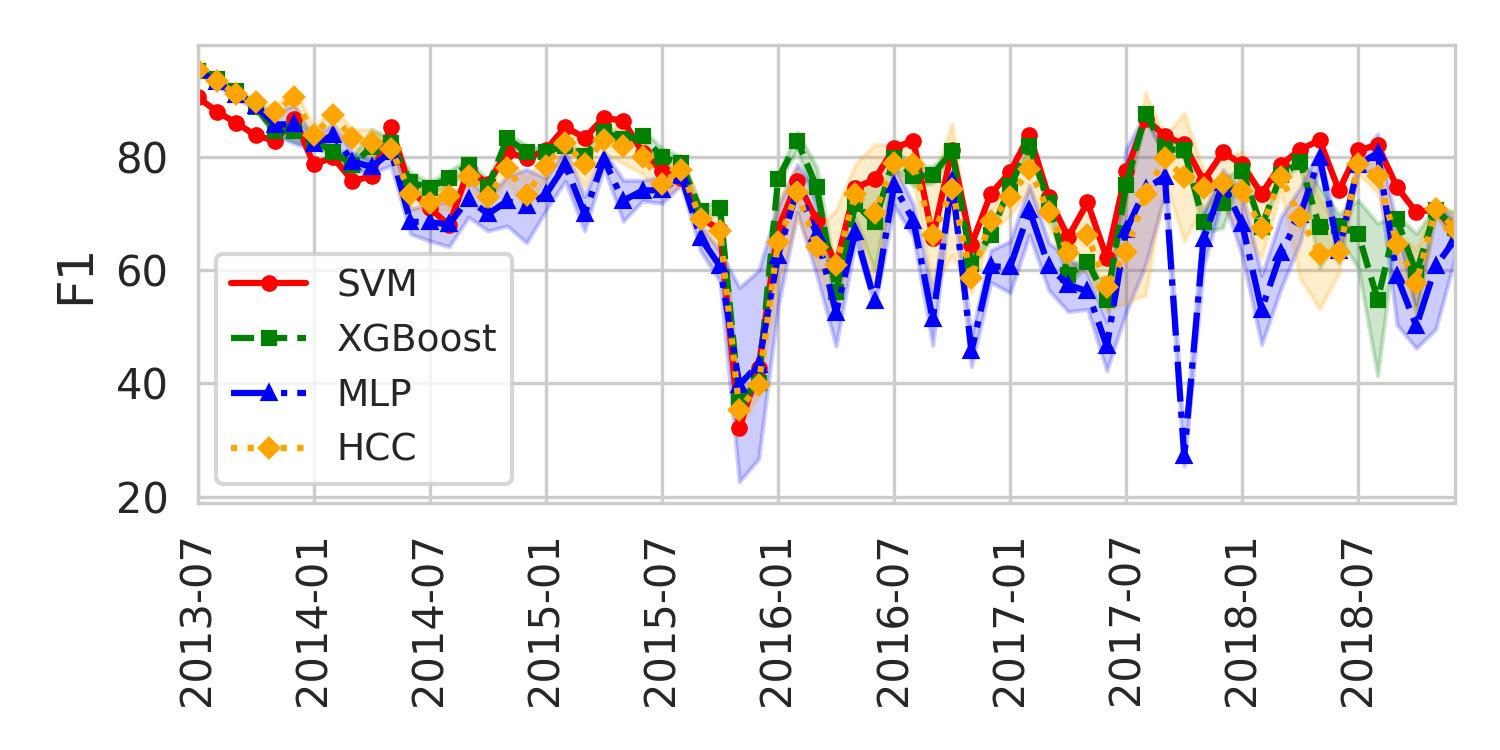}
    \caption{F1-score over the test months on the APIGraph for (left) deduplicated (right) duplicated datasets}
    \label{fig:var-apigraph}
\end{figure*}

\begin{figure*}
\centering
  \begin{subfigure}{0.32\textwidth}
    \includegraphics[width=\linewidth]{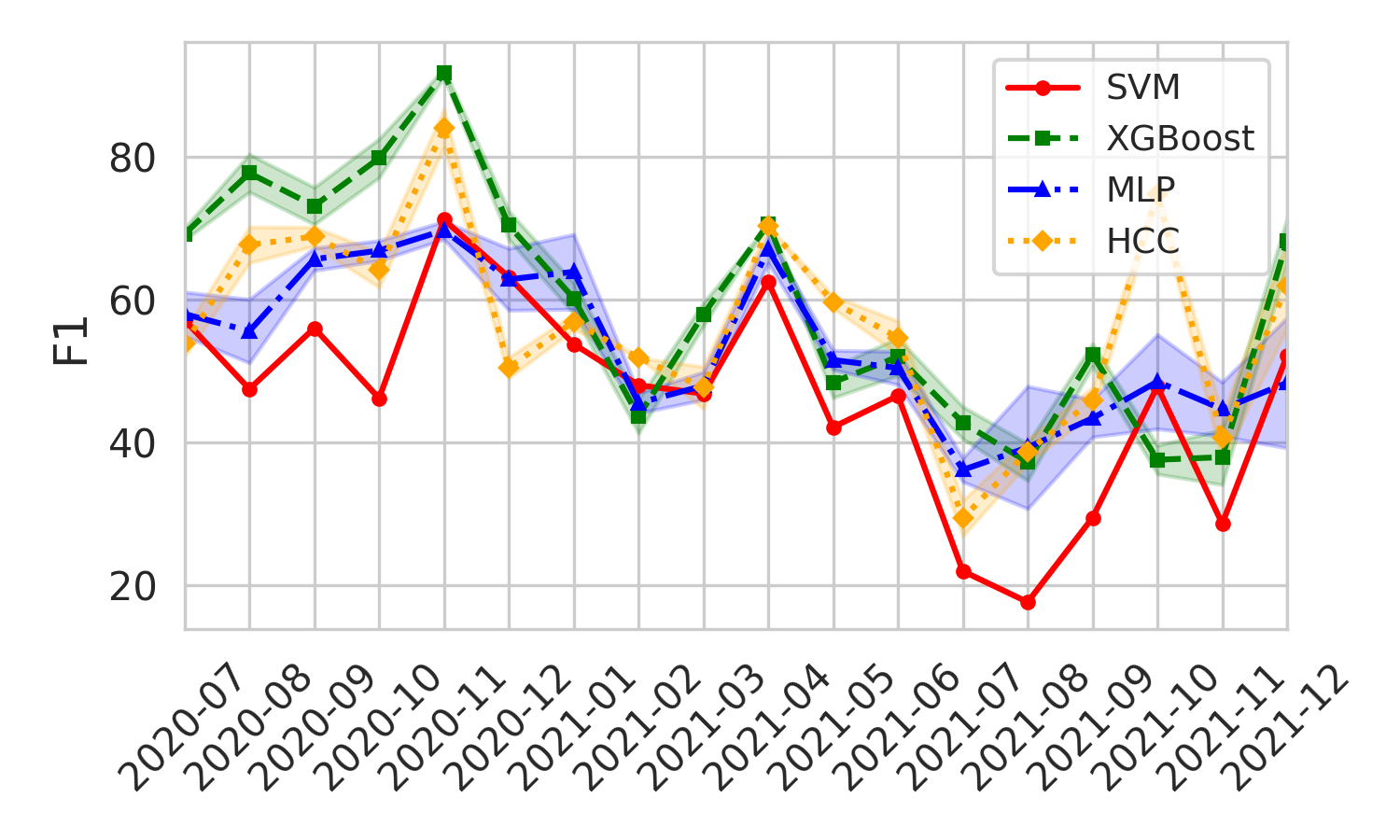}
    \caption{F1-score on deduplicated data}
  \end{subfigure}
  \begin{subfigure}{0.32\textwidth}
    \includegraphics[width=\linewidth]{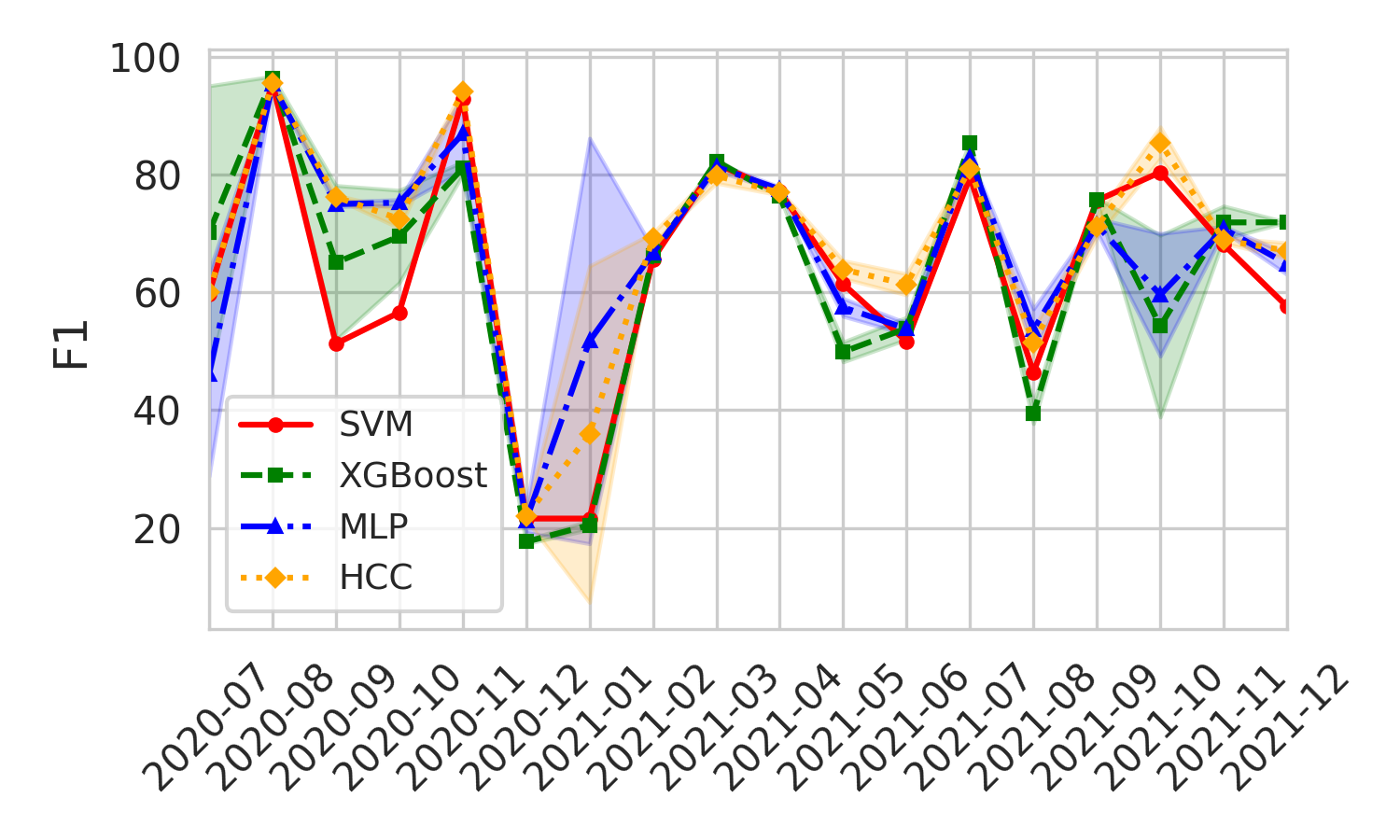}
    \caption{F1-score on duplicated dataset}
  \end{subfigure}
  \begin{subfigure}{0.32\textwidth}
    \includegraphics[width=\linewidth]{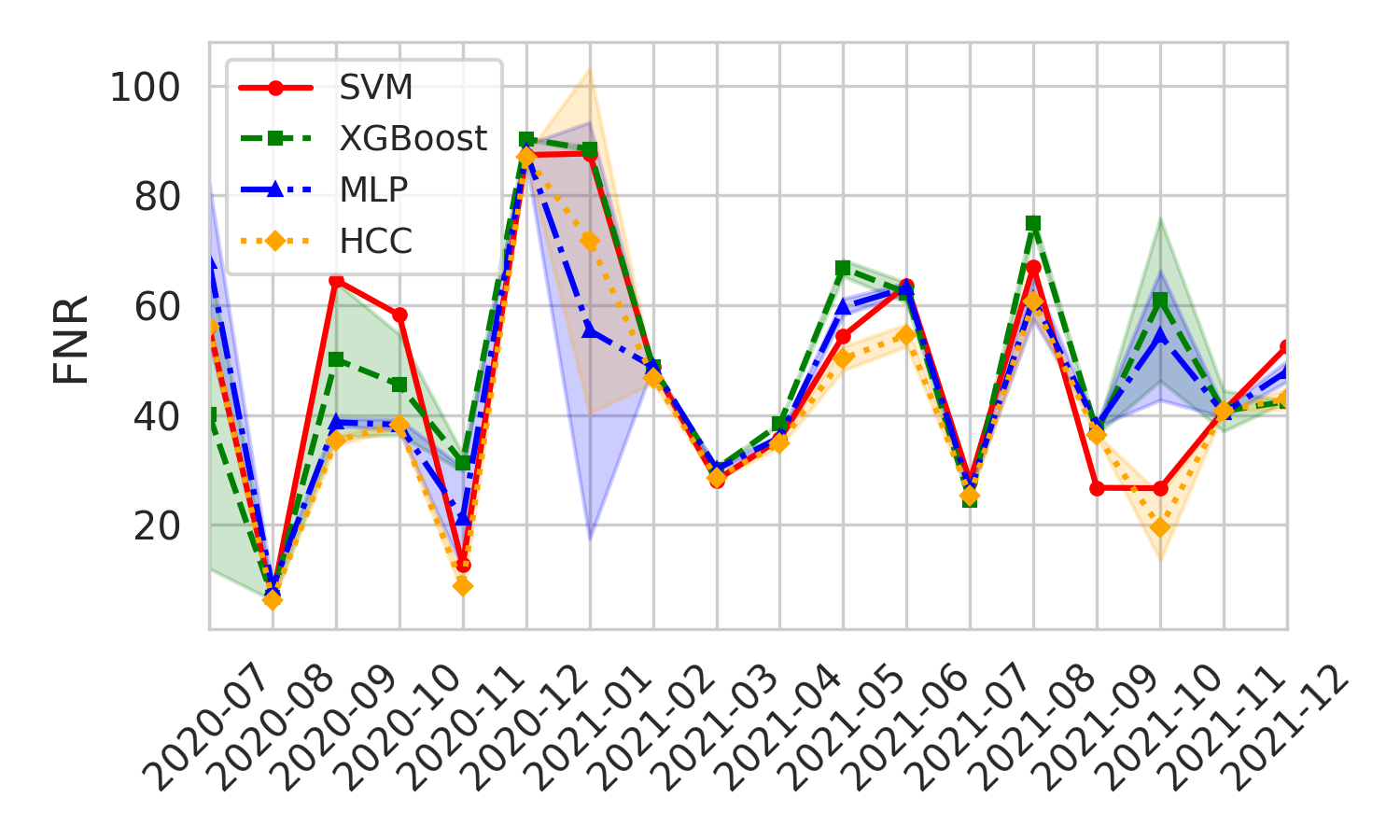}
    \caption{FNR on duplicated dataset}
  \end{subfigure}
  
  \caption{Performance in test months on the Drebin datasets for Merged Training setting}
  \label{fig:var-drebin}
\end{figure*}

\section{Offline Learning Results \& Analysis} \label{sec-offline-res}

\subsection{Malware Detection Performance}

We present the results of six models in Table~\ref{tab:res-static-drebin} for the Drebin dataset and Table~\ref{tab:res-static-apigraph} for the APIGraph dataset. These results encompass the various experimental settings discussed previously. Key findings are summarized as follows:

\begin{enumerate}
    \item \textbf{Impact of duplicates on model performance:} Deduplicated datasets generally perform worse than those with duplicates, indicating that duplicates may inflate performance metrics. For instance, in the Drebin dataset's merged training setting, the F1-score differences for the SVM and SCC models are 17.0 and 17.3, respectively, between duplicate and deduplicated datasets.

    \item \textbf{Benefits of merged training:} Merging validation data with training data before final evaluation consistently enhances performance. In the Drebin deduplicated dataset, the average F1-score improvement ranges from 8.3 (RF) to 20.9 (HCC), demonstrating the benefits of additional training data and its alignment with practical applications.

    \item \textbf{Duplicates influence model preference:} Duplicates can skew model comparisons. For example, in the Drebin dataset merge-training setting, SCC outperforms XGBoost by 11 in average F1-score. However, after deduplication, XGBoost outperforms SCC by 2.1 in average F1-score.

    \item \textbf{XGBoost as a strong baseline on deduplicated datasets:} 
With proper tuning, XGBoost achieves the highest average F1-score on the deduplicated Drebin dataset and remains competitive on APIGraph, highlighting its effectiveness as a baseline model. In contrast, while SVM performs well on APIGraph, its performance drops significantly on the Drebin dataset after deduplication.

\end{enumerate}

\begin{figure*}
  \centering  \includegraphics[width=0.45\textwidth]{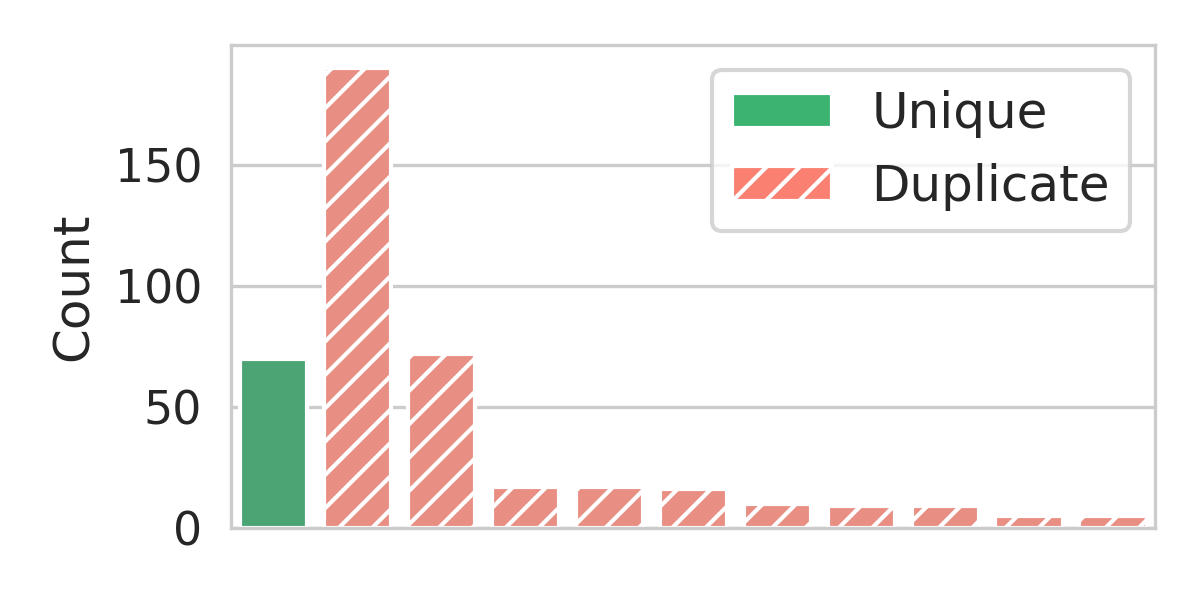}    \includegraphics[width=0.45\textwidth]{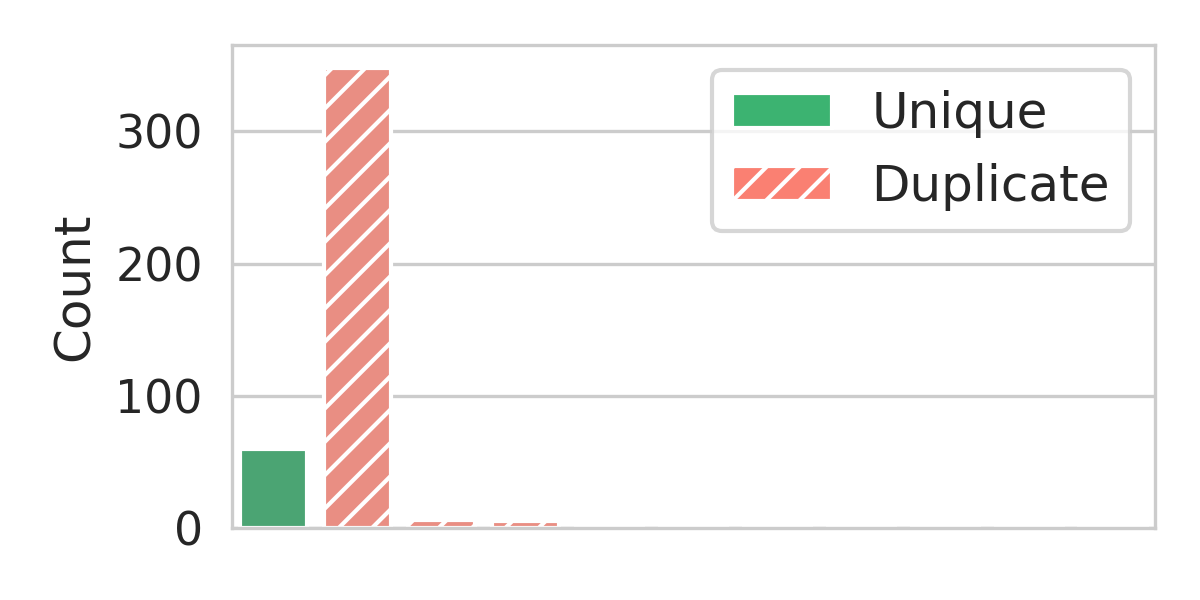}
    \caption{Frequency of unique and duplicate malware samples. Left: APIGraph in November 2015 with 190 of 488 samples sharing identical features. Right: Drebin in January 2021, with 348 of 385 samples duplicates of the same input.}

    \label{fig:dupe-sample}
\end{figure*}

\subsection{Duplicates and Variance in Performance} \label{sec-offline-variance}

Figures~\ref{fig:var-apigraph} and \ref{fig:var-drebin} show the month-by-month performance of four models (SVM, XGBoost, MLP, HCC) on the APIGraph and Drebin datasets under the merged training setting for both duplicated and deduplicated datasets. Key conclusions from the results are:

\begin{enumerate}
    \item \textbf{Impact of duplicates on performance trends:} On both datasets, models generally exhibit a downward performance trend over time due to concept drift. However, this trend is less evident in the original datasets with duplicates, where sudden fluctuations occur. For example, all models experienced a performance drop in November 2015 on the APIGraph dataset, explained by duplicate inputs, as shown in Figure~\ref{fig:dupe-sample} (left). Of the 488 malware samples that month, 190 have identical input features. If a model fails to detect one correctly, it fails for all, resulting in a significant performance drop.

    \item \textbf{High variance due to duplicates:} As seen in Figure~\ref{fig:var-drebin}(b), duplicates can cause high variance in model performance. Figure~\ref{fig:var-drebin}(c) shows that this effect is primarily due to variance introduced by false negative rates (FNRs). Figure~\ref{fig:dupe-sample}(right) illustrates the frequency of unique, and top-10 duplicated malware samples for January 2021 on the Drebin dataset. Specifically, 348 out of 385 samples are duplicates of the same input feature, resulting in significant variance for models relying on random initialization, such as neural networks. Two of five MLPs trained with different seeds achieved high F1 scores (93.7 and 94.0), while the others scored much lower (21.4, 24.6, 24.8), indicating high fluctuations due to duplicates. A theoretical classifier correctly identifying all malware samples this month has an FNR of 0. If it fails to detect this single duplicated sample, the FNR rises to 90.4\%. Conversely, a classifier detecting no samples has an FNR of 100\%, but identifying only the duplicated sample reduces the FNR to 9.6\%. This highlights the need to deduplicate datasets for models relying on random initialization to prevent performance disparities caused by different initializations.

      \item \textbf{Importance of reporting performance with multiple
seeds:} Although deduplication reduces extreme variance in model performance for certain months, neural networks can still show performance variation due to random initialization, as seen in Figure~\ref{fig:var-drebin}(a). For instance, in four test months (2021-01, 2021-08, 2021-10, 2021-12), the MLP had a standard deviation of average F1-scores greater than 5, even on the deduplicated Drebin dataset. Therefore, accounting for this randomness when reporting neural network performance is essential.
\end{enumerate}

\subsection{Effective Model Selection for Robust Baselines} \label{sec-hpotxsboost}

\begin{figure}[h]
    \centering
    \includegraphics[width=0.49\textwidth]{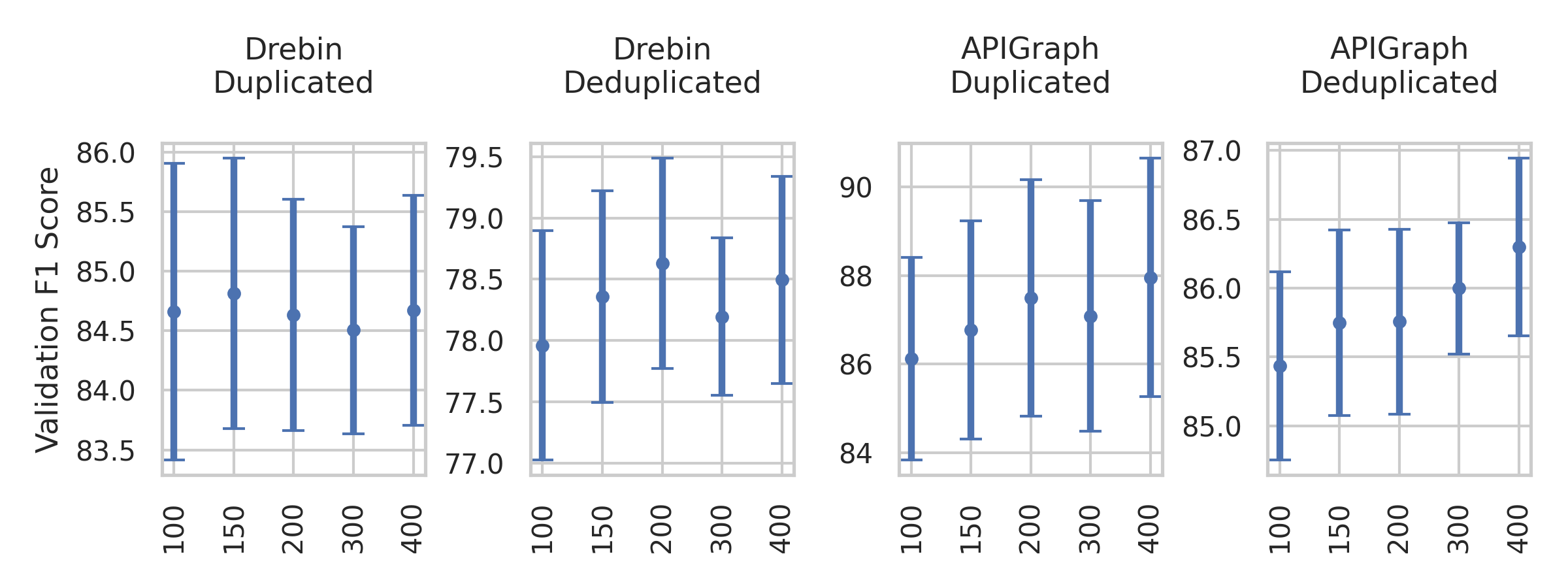}
    \caption{Effect of number of boosting rounds during learning for XGBoost model on different datasets}
    \label{fig:xgboost-round}
\end{figure}

Our offline learning experiments show that XGBoost consistently outperforms other models across settings, differing from the results in~\cite{chen2023continuous} primarily due to an expanded hyperparameter search space. We included additional parameters and wider ranges (see Appendix~\ref{appendix-hpo}), with the number of boosting rounds being particularly important. While~\cite{chen2023continuous} tested only 10–100 rounds, we found that up to 400 rounds significantly improved performance, especially on the APIGraph dataset. This highlights the importance of thorough hyperparameter tuning for establishing strong baselines~\cite{arp2022and}. Figure~\ref{fig:xgboost-round} shows the impact of boosting rounds on validation performance across datasets.

\begin{table*}[]
\caption{Performance of different models on the Drebin Dataset for continuous active learning with 50 \& 100 annotation budget per month}
\centering
\resizebox{\textwidth}{!}{%
\setlength{\tabcolsep}{3.25pt}
\begin{tabular}{l|ccc|ccc|ccc|ccc} 
\toprule
\multirow{2}{*}{\textbf{Model}} & \multicolumn{6}{c|}{\textbf{Budget=50}} & \multicolumn{6}{c}{\textbf{Budget=100}} \\
\cmidrule(lr){2-7} \cmidrule(lr){8-13}
& \multicolumn{3}{c|}{\textbf{Merged Training}} & \multicolumn{3}{c|}{\textbf{Holdout Training}} & \multicolumn{3}{c|}{\textbf{Merged Training}} & \multicolumn{3}{c}{\textbf{Holdout Training}} \\
\cmidrule(lr){2-4} \cmidrule(lr){5-7} \cmidrule(lr){8-10} \cmidrule(lr){11-13}
& F1 & FPR & FNR & F1 & FPR & FNR & F1 & FPR & FNR & F1 & FPR & FNR \\
\midrule
RF & 61.9±0.73&\underline{0.13±0.02}&53.0±0.88&62.7±1.11&\underline{0.11±0.00}&52.4±1.11&62.8±1.24&\underline{0.13±0.01}&52.1±1.30&63.6±0.47&\underline{0.12±0.03}&51.3±0.40 \\ 
SVM &64.4±0.00&0.22±0.00&48.7±0.00&59.6±0.00&0.23±0.00&54.1±0.00&68.1±0.00&0.24±0.00&44.6±0.00&63.9±0.00&0.26±0.00&49.1±0.00 \\ 
XGBoost & \underline{82.2±0.60}&0.15±0.00&\underline{27.3±0.76}&\underline{81.1±0.15}&0.14±0.01&\underline{29.2±0.21}&\underline{82.5±0.29}&0.15±0.01&\underline{26.9±0.32}&\underline{82.5±0.45}&0.14±0.01&\underline{27.0±0.57} \\ 
MLP & 67.9±0.65&0.49±0.03&43.3±0.70&61.8±0.54&0.35±0.02&51.0±0.49&70.2±0.42&0.38±0.03&40.6±0.46&65.7±3.37&0.36±0.06&46.2±4.17 \\ 
SCC & 71.0±0.64&0.59±0.04&37.4±0.29&68.2±1.83&0.50±0.07&41.6±2.21&73.7±1.20&0.57±0.10&33.7±1.35&72.1±0.88&0.48±0.07&37.0±1.44 \\ 
HCC & 73.5±0.27 & 0.66±0.04 & 32.8±0.85 & 68.5±3.38&0.57±0.05&40.3±4.37&75.6±0.96&0.68±0.06&29.4±1.07&72.6±2.10&0.57±0.02&35.2±2.64 \\ 
\bottomrule
\end{tabular}
}

\label{tab:res-al-drebin}
\end{table*}

\begin{table*}[]
\caption{Performance of different models on the APIGraph Dataset for continuous active learning with 50 \& 100 annotation budget per month}
\centering
\resizebox{\textwidth}{!}{%
\setlength{\tabcolsep}{3.25pt}
\begin{tabular}{l|ccc|ccc|ccc|ccc} 
\toprule
\multirow{2}{*}{\textbf{Model}} & \multicolumn{6}{c|}{\textbf{Budget=50}} & \multicolumn{6}{c}{\textbf{Budget=100}} \\
\cmidrule(lr){2-7} \cmidrule(lr){8-13}
& \multicolumn{3}{c|}{\textbf{Merged Training}} & \multicolumn{3}{c|}{\textbf{Holdout Training}} & \multicolumn{3}{c|}{\textbf{Merged Training}} & \multicolumn{3}{c}{\textbf{Holdout Training}} \\
\cmidrule(lr){2-4} \cmidrule(lr){5-7} \cmidrule(lr){8-10} \cmidrule(lr){11-13}
& F1 & FPR & FNR & F1 & FPR & FNR & F1 & FPR & FNR & F1 & FPR & FNR \\
\midrule
RF & 87.1±0.15&\underline{0.27±0.00}&19.4±0.28&87.4±0.20&\underline{0.30±0.01}&18.6±0.34&89.0±0.11&\underline{0.26±0.01}&16.4±0.13&89.4±0.11&\underline{0.28±0.01}&15.6±0.11 \\ 
SVM &83.7±0.00&0.60±0.00&21.4±0.00&84.4±0.00&0.67±0.00&19.2±0.00&84.8±0.00&0.63±0.00&19.0±0.00&85.2±0.00&0.75±0.00&16.9±0.00 \\ 
XGBoost & \underline{88.7±0.21}&0.55±0.01&\underline{13.8±0.31}&\underline{88.9±0.11}&0.53±0.03&\underline{13.7±0.11}&\underline{90.4±0.13}&0.48±0.01&\underline{11.6±0.21}&\underline{90.6±0.18}&0.45±0.01&\underline{11.5±0.16} \\ 
MLP & 85.7±0.67&0.47±0.03&19.6±1.12&85.7±0.40&0.47±0.05&19.5±1.10&87.4±0.29&0.47±0.02&16.6±0.49&87.2±0.09&0.49±0.04&16.6±0.60 \\ 
SCC & 85.6±0.37&0.85±0.07&15.1±0.27&86.0±0.71&0.81±0.08&14.8±0.31&86.1±0.38&0.95±0.07&12.8±0.61&85.7±2.80&2.03±2.51&12.9±0.48 \\ 
HCC & 87.1±0.36&0.64±0.05&15.6±0.95&86.7±0.15&0.56±0.03&17.2±0.36&87.5±0.78&0.56±0.03&15.5±1.11&87.4±0.27&0.59±0.02&15.6±0.14 \\ 
\bottomrule
\end{tabular}
}

\label{tab:res-al-apigraph}
\end{table*}

\subsection{Duplicates \& Model Selection}
\begin{figure}[t]
    \centering
    \includegraphics[width=0.45\textwidth]{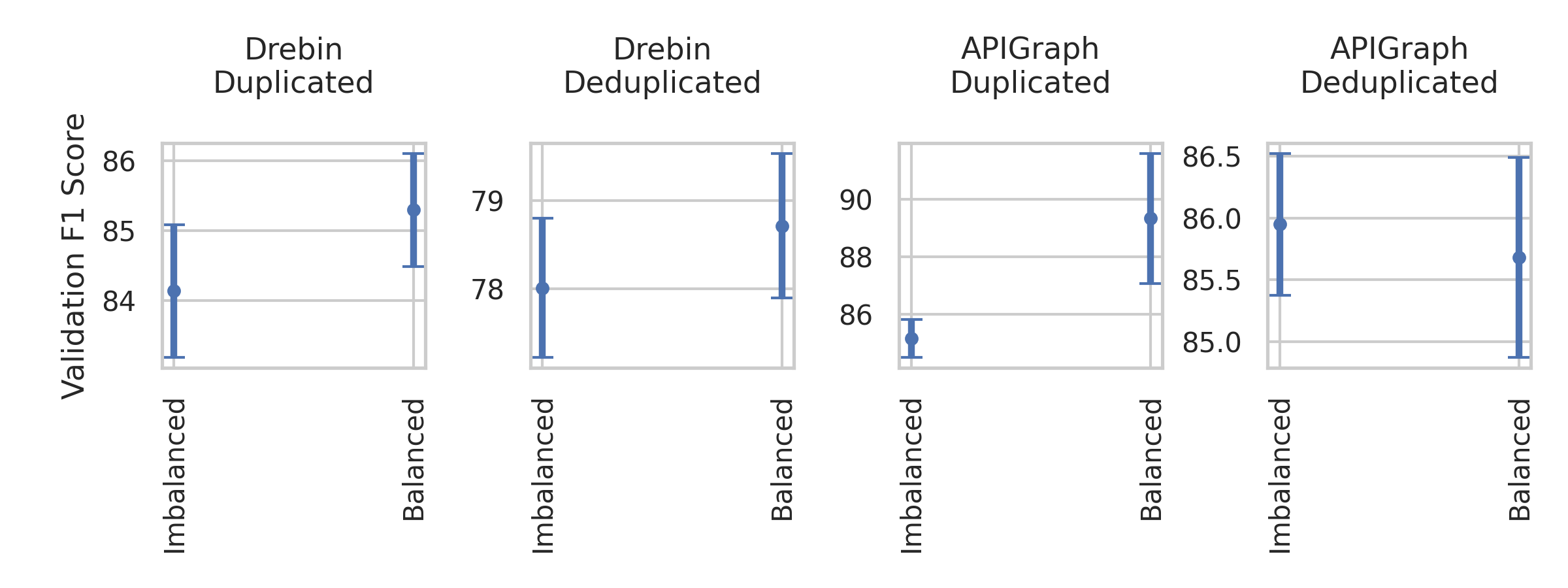}
    \caption{Effect of balancing the weight of the classes during learning for XGBoost model on different datasets}
    \label{fig:xgboost-balance}
\end{figure}

The presence of duplicates can affect hyperparameter tuning and model selection. As shown in Figure~\ref{fig:xgboost-balance}, adjusting the \texttt{scale\_pos\_weight} parameter to balance classes improves validation performance for duplicated datasets but can reduce it on the deduplicated APIGraph dataset, indicating that duplicates influence optimal hyperparameter behavior.

\begin{figure}
    \centering
    \includegraphics[width=0.48\textwidth]{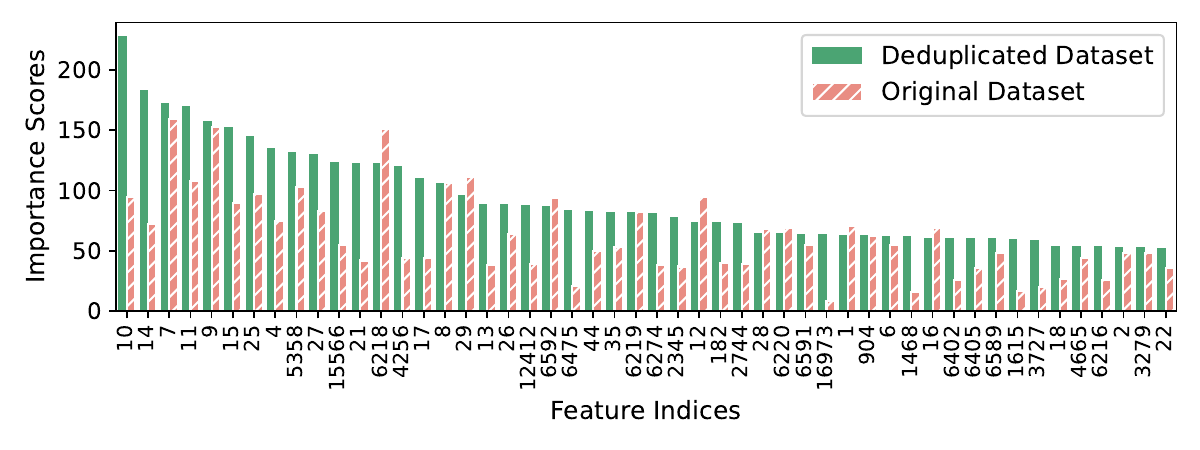}
    \caption{Comparison of feature weights for the top 50 features in XGBoost models trained on the deduplicated Drebin dataset and their corresponding weights in the model trained on the original duplicated dataset.}
    \label{fig:xgboost-xai-dedup2org}
\end{figure}

Figure~\ref{fig:xgboost-xai-dedup2org} further shows that duplicates alter feature weighting in XGBoost. Comparing models trained on deduplicated and original Drebin datasets, the top 50 features from the deduplicated model display markedly different importance distributions when evaluated on the duplicated data (feature names in Appendix~\ref{app:mapping}).

\section{Continuous Active Learning Results \& Analysis} \label{sec-active-res}

In the continuous active learning setting, duplicates pose additional challenges beyond those in offline learning. Since active learning relies on human experts to select samples for annotation, increasing the diversity of selected samples is crucial~\cite{settles2009active}. Duplicates can lead to redundant selections, wasting the annotation budget. A simple heuristic can identify previously annotated samples and replace them with their existing annotations. Thus, using duplicated datasets in an active learning setting is impractical, and our analysis primarily focuses on deduplicated datasets.

We present the results of the six models in Tables~\ref{tab:res-al-drebin} and \ref{tab:res-al-apigraph} for the Drebin and APIGraph datasets, discussing merged and holdout training settings. Key findings are summarized below:

\begin{enumerate}
    \item \textbf{XGBoost as a strong baseline in continuous active learning:} XGBoost achieves the highest average F1 score across all experimental settings in continuous active learning. This is especially evident in the challenging Drebin dataset, where it surpasses the second-best model by 8.7\% in the merged training setting. This highlights the importance of selecting a strong baseline model and underscores XGBoost's effectiveness in Android malware detection.

    \item \textbf{Benefits of Contrastive Learning for Neural Networks:} The neural networks trained with contrastive learning, SCC, and HCC outperform baseline neural networks in all settings, indicating the effectiveness of contrastive learning in enhancing robustness to concept drift. HCC further outperforms SCC, demonstrating the advantages of hierarchical contrastive learning and the pseudo-loss sample selector introduced in~\cite{chen2023continuous}. However, more improvements are needed to match XGBoost's performance.

    \item \textbf{Random Forest as a Competitive Baseline on APIGraph:} Even simpler tree-based methods like Random Forest achieve competitive performance on the APIGraph dataset, often surpassing more complex neural networks. With consistently low false positives, RF is a strong candidate model for this dataset, illustrating that simpler models can compete effectively with complex ones in a compact feature space.

    \item \textbf{Performance Gains from Merged Training:} Merging the validation set with the training set can enhance performance on test data, particularly for neural networks. However, the performance difference is less pronounced than offline learning since models are retrained during validation months with a subset of samples.
\end{enumerate}

\begin{figure}
    \centering
    \includegraphics[width=0.8\linewidth]{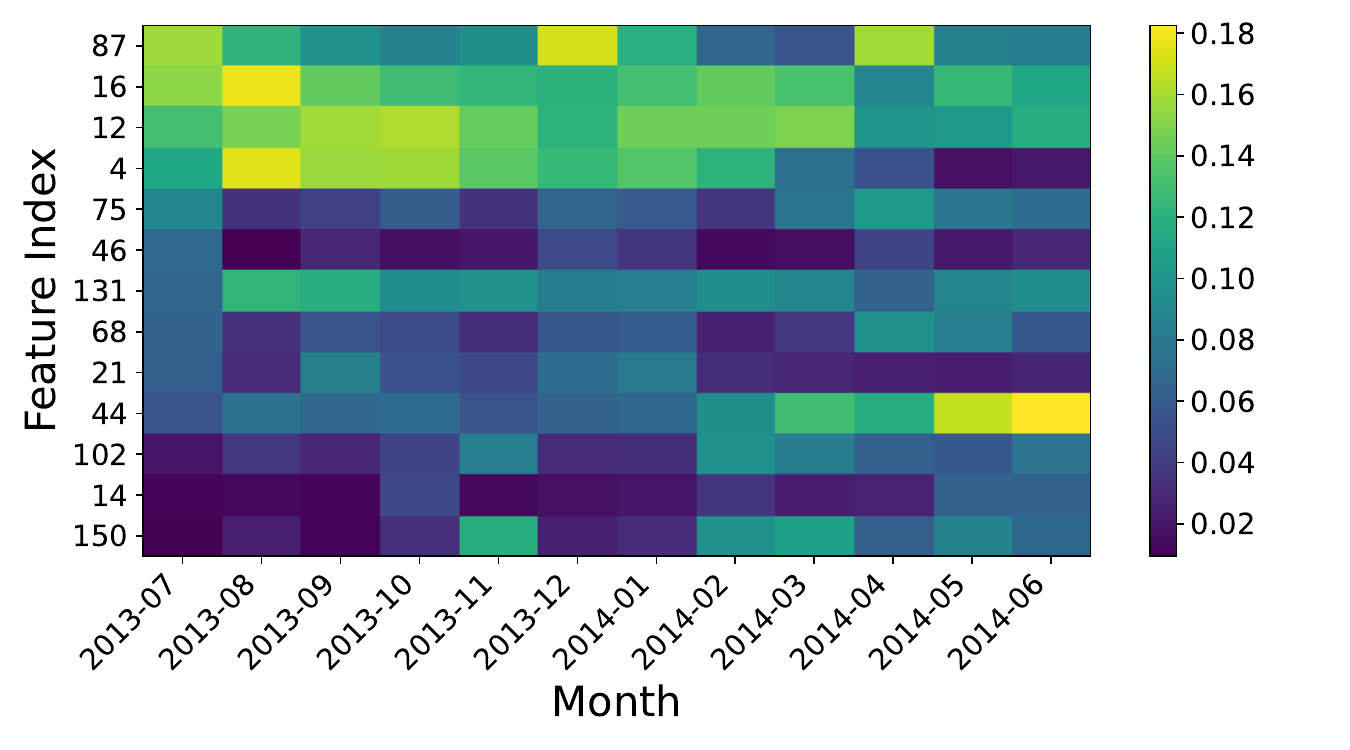}
    \caption{Evolution of normalized XGBoost feature importance from July 2013 to June 2014.}

    \label{fig:shapxgboost}
\end{figure}

\begin{figure}[t]
    \centering
    \includegraphics[width=0.8\linewidth]{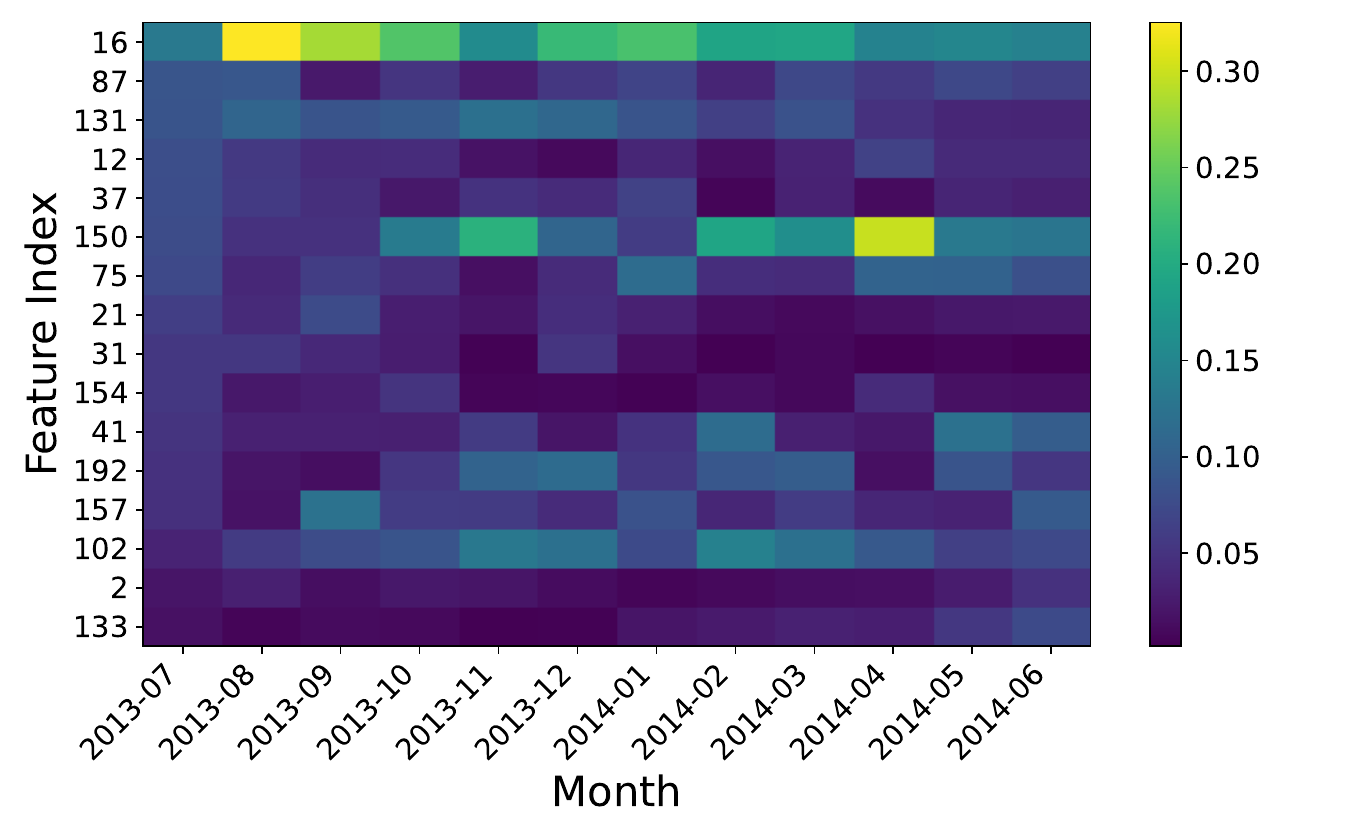}
    \caption{Evolution of feature importance for MLP models from July 2013 to June 2014.}

    \label{fig:shapmlp}
\end{figure}

\begin{figure}[t]
    \centering
    \includegraphics[width=0.8\linewidth]{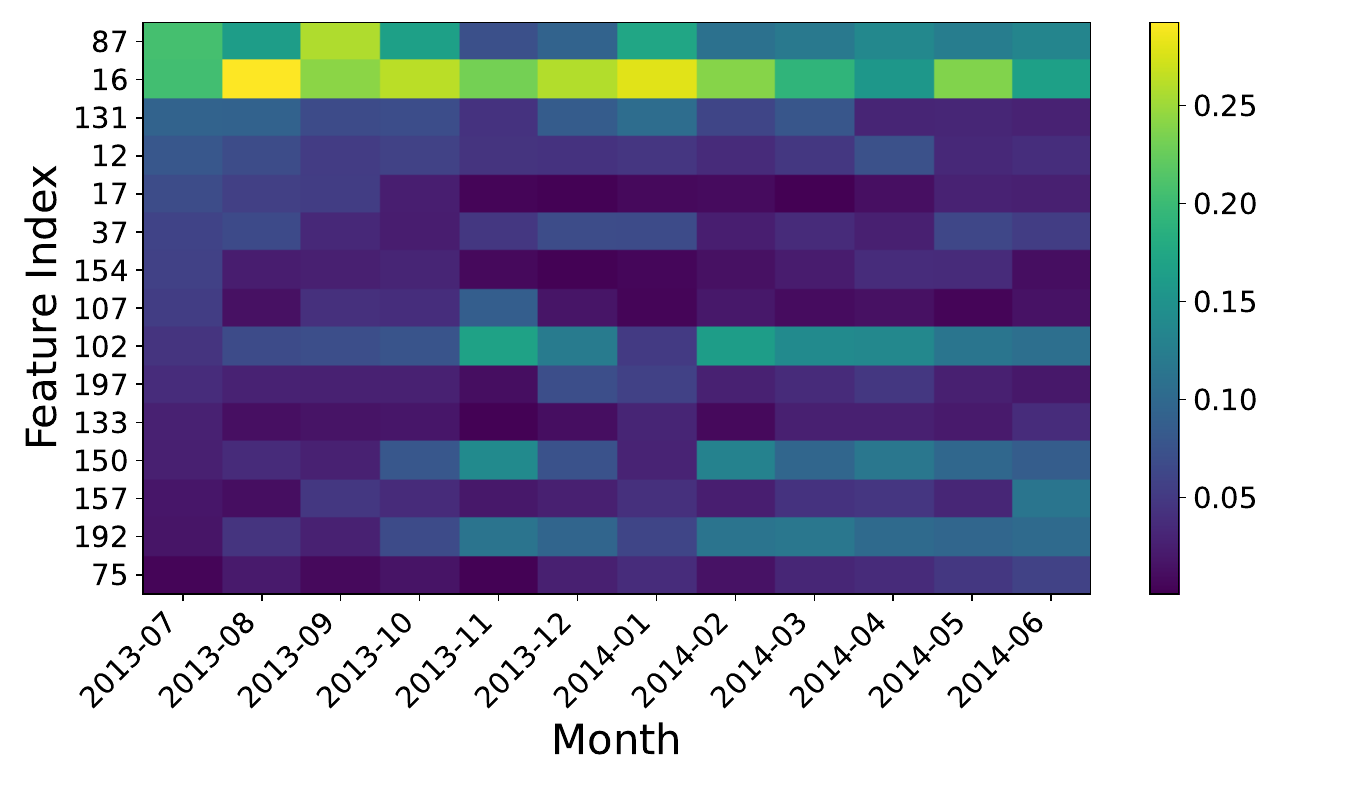}
    \caption{Evolution of normalized feature importance for HCC models from July 2013 to June 2014.}

    \label{fig:shaphcc}
\end{figure}

\section{Post-hoc explainability}

To examine how learned features change over time, we analyzed XGBoost, MLP, and HCC classifiers on the APIGraph dataset using SHAP-based explanations~\cite{lundberg2017unified}. For each model, we constructed a reference set by taking the union of the top-10 most important features identified in the first and last months of the initial one-year period (July 2013–June 2014), using TreeSHAP for XGBoost and KernelSHAP for MLP and HCC. We then fixed this set as a basis for comparison and, for each subsequent month, recomputed feature importances using correctly classified malware samples from that month’s test data. All importances were normalized to ensure comparability. The resulting heatmaps (Figures~\ref{fig:shapxgboost}, \ref{fig:shapmlp}, and \ref{fig:shaphcc}) reveal the temporal dynamics of feature relevance across models.

Figure~\ref{fig:shapxgboost} shows that in XGBoost models several features (e.g., Feature~12 [\texttt{GetDeviceID}], Feature~16 [\texttt{GetSubscriberID}], and Feature~87 [\texttt{SendSMS}]) remain consistently influential across months. In contrast, Feature~4 [\texttt{ReadPhoneStateUsed}] is prominent in the early months but declines in importance toward the end of the year, where Feature~44 [\texttt{ReadPhoneStateRequested}] gains greater weight.  

For neural network models, the importance distribution is more static, with less variation across months. Despite their different training strategies (HCC being trained with hierarchical contrastive features), MLP and HCC show substantial overlap in their top features (e.g., the top four features are identical in the first month). While MLP consistently assigns high importance to Feature~16 [\texttt{GetSubscriberID}], HCC emphasizes both Feature~16 [\texttt{GetSubscriberID}] and Feature~87 [\texttt{SendSMS}] as particularly influential.


\begin{figure}[t]
    \centering
    \includegraphics[width=0.4\linewidth]{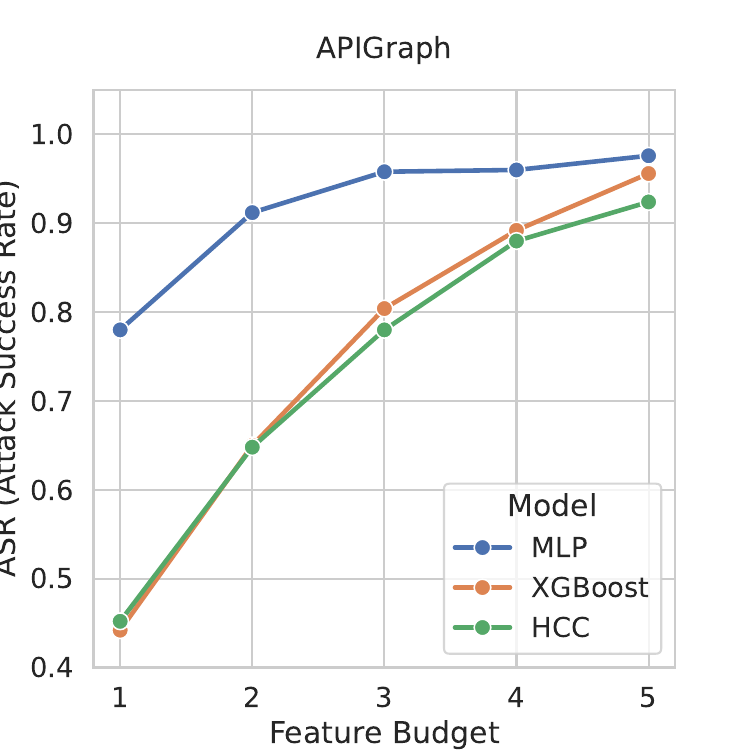}
    \hfill
    \includegraphics[width=0.4\linewidth]{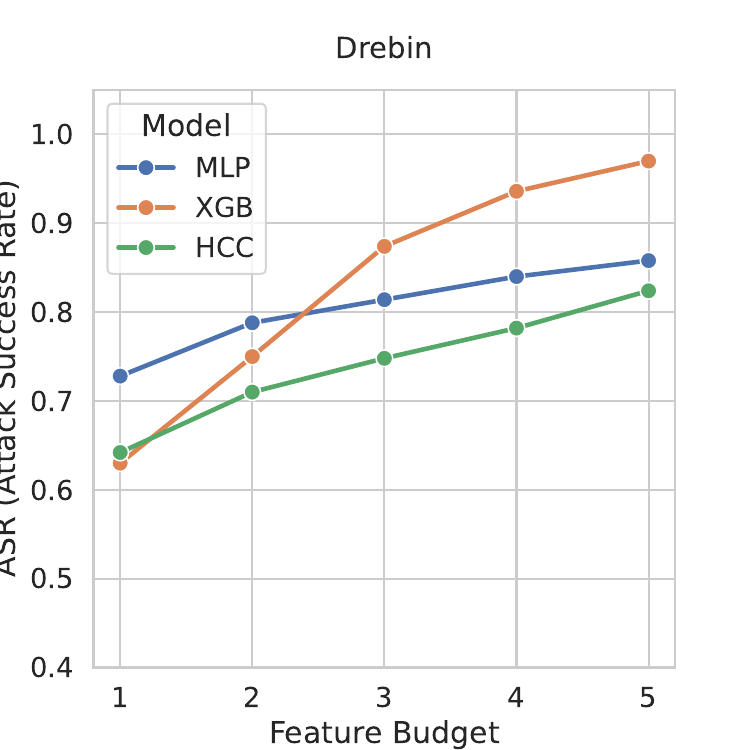}
    \caption{Comparison of attack success rate}
    \label{fig:asr}
\end{figure}

\section{Adversarial Robustness}
We evaluate a pure black-box, score-based, feature-space evasion~\cite{simonetto2024constrained} on APIGraph and Drebin datasets. The Constrained Adaptive Attack (CAA) is a state-of-the-art method for adversarial evasion on tabular data that combines a fast gradient-based phase (CAPGD) with a stronger search phase. In our setting, we disable CAPGD and use only the search phase: a best-first beam search over single-bit add-only flips that batch-queries model scores, retains the lowest-score successors in a fixed-width beam (width 16; node cap 64), and halts upon reaching the benign threshold, exhausting the flip budget, or hitting the node cap. The attacker may flip up to \(B\) binary features per sample under an add-only constraint; all features are treated as mutable. We attack 500 randomly sampled malware test points per model, while benign samples remain unchanged.

Results are shown in Figure~\ref{fig:asr} for budgets \(B \in \{1,2,3,4,5\}\). Even with small budgets, the attack achieves high success across all models. On APIGraph, XGBoost and HCC are more robust than MLP, whereas on Drebin, XGBoost shows higher ASR than MLP for larger budgets (\(B=3,4,5\)), indicating that higher clean accuracy does not necessarily imply stronger adversarial robustness.
\section{Limitations \& Discussions}
\label{sec:limitations}

\begin{figure}[t]
    \centering
    \includegraphics[width=0.38\textwidth]{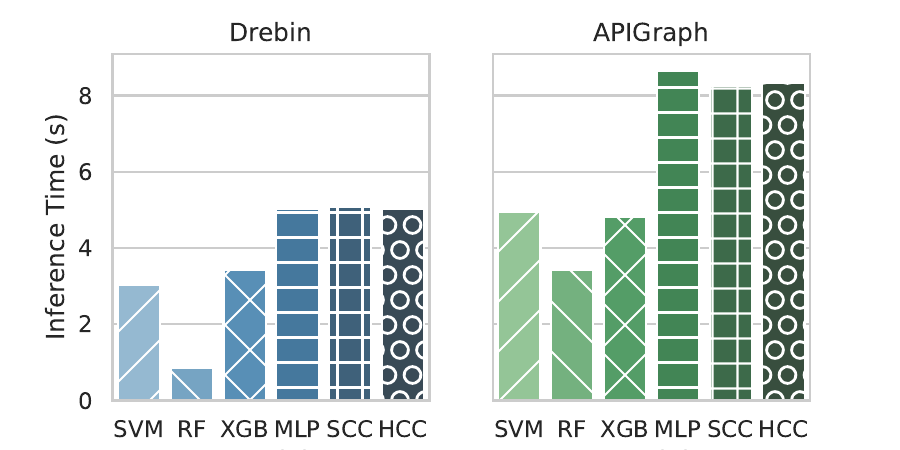}
    \caption{Inference time (in seconds) of different machine learning models on the Drebin (left) and APIGraph (right).}
    \label{fig:time}
\end{figure}

\textbf{Computation Time.}
Figure~\ref{fig:time} illustrates a comparison of the aggregate inference times for various machine learning models evaluated on the Drebin and APIGraph test datasets. The results show that neural models consistently incur higher inference latency compared to traditional models such as Random Forest (RF) and Support Vector Machine (SVM). The experiments were conducted using Python~3.11.6 on Red Hat Enterprise Linux~9.4, running on an Intel Xeon Gold CPU at 2.20\,GHz.

\begin{figure}[t]
    \centering
    \includegraphics[width=0.8\columnwidth]{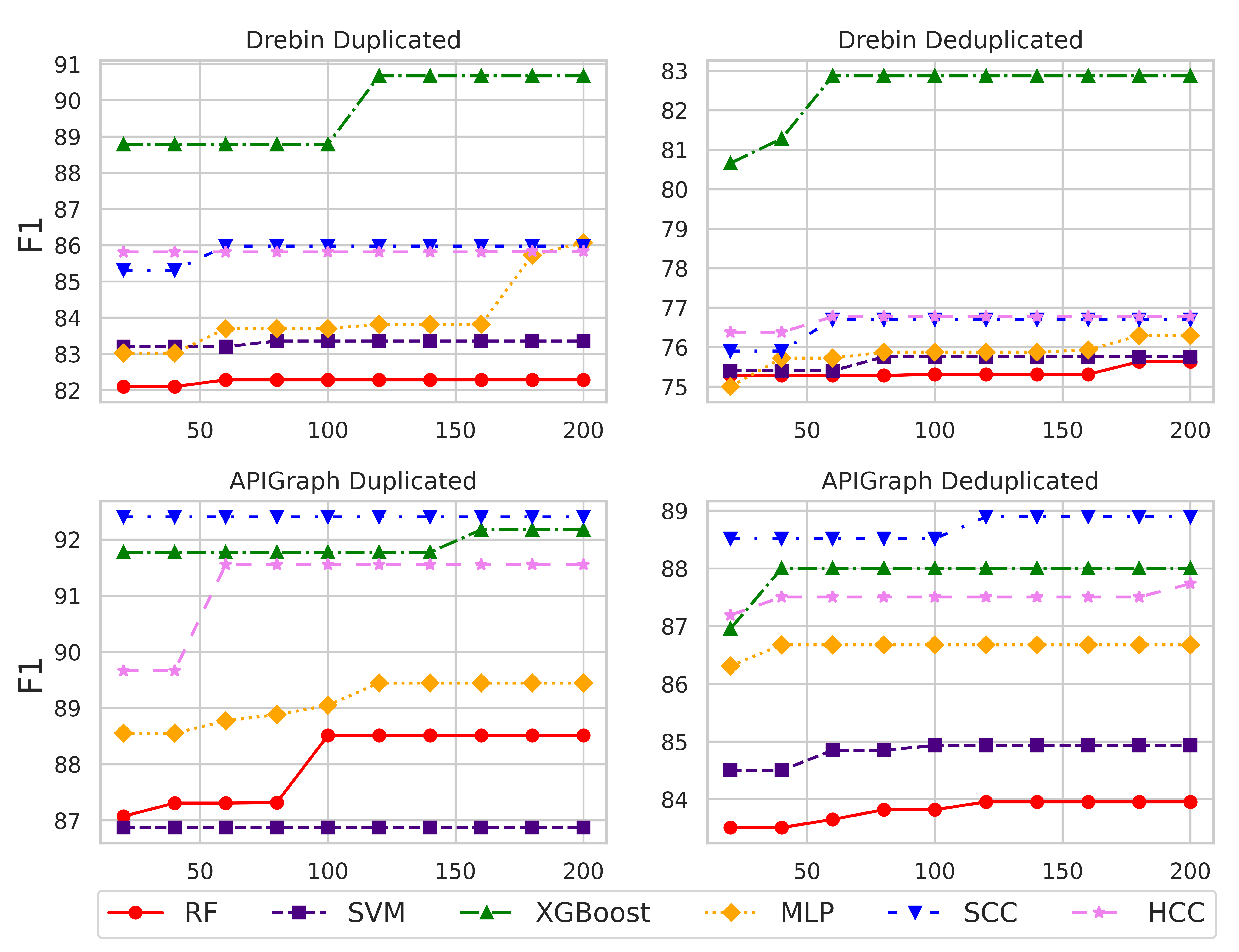} 

    \caption{Performance on validation set as number of hyperparameter trials is increased for offline learning}
    \label{fig:offline-hpo}
\end{figure}

\begin{figure}[t]
    \centering
    \includegraphics[width=0.8\columnwidth]{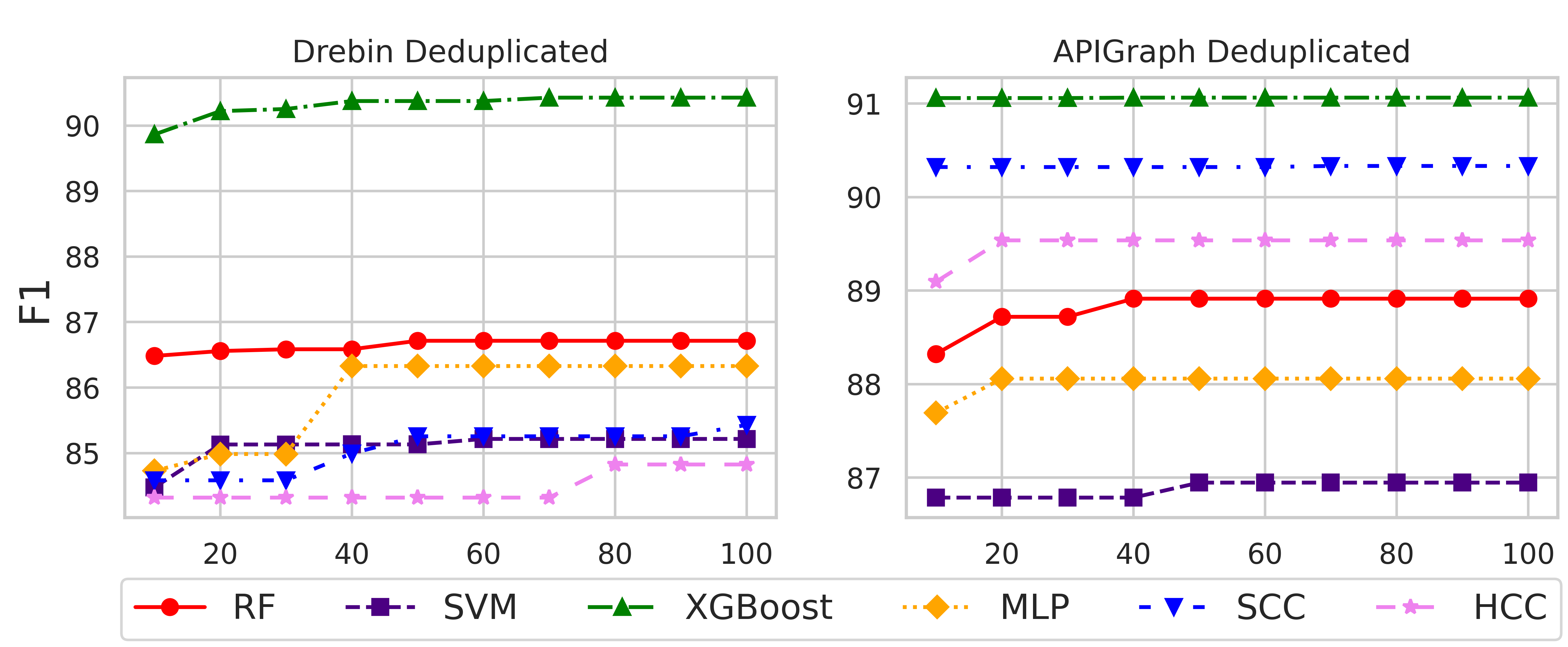} 

    \caption{Validation performance versus number of hyperparameter trials in continuous active learning.}

    \label{fig:al-hpo}
\end{figure}

\textbf{Number of Trials and Performance Convergence.} We use random search for hyperparameter tuning, following standard practice in domain generalization where training and test sets are not i.i.d.~\cite{domainbed,yu2024rethinking}. This approach ensures fair comparison without biasing toward models with larger search spaces. To maintain rigor, we allocate sufficient trials for stable validation results. As shown in Figures~\ref{fig:offline-hpo} and~\ref{fig:al-hpo}, most models converge within 100 trials for offline learning and even fewer in continuous learning, indicating limited sensitivity to initial hyperparameters. While validation performance generally reflects test performance, exceptions exist, such as the RF model on the Drebin dataset.



\textbf{Annotation Interval.}  
We use a fixed interval for continuous active learning experiments, as widely adopted in prior work on Android malware detection~\cite{pendlebury2019tesseract,jordaney2017transcend,barbero2022transcending,kan2021investigating,xu2019droidevolver,chen2023continuous}. In practical deployments, concept drift detection could decide when to retrain the model~\cite{jordaney2017transcend}, but this complicates comparisons across active learning methods since different models may trigger retraining at different times. Moreover, altering the retraining interval while keeping the annotation budget fixed can impact results; for example, increasing the interval to two months may reduce performance because the model is evaluated on a longer horizon of unseen data before retraining, compared to the current one-month setup.

\textbf{Annotation Budget in the Presence of Duplicates.}  
In our active learning setup, we assume a fixed annotation budget. For practical deployment, however, duplicates introduce an additional challenge: the annotation process should prioritize uncertain samples while ensuring that no sample is annotated more than once, as repeated annotations waste valuable resources. While this constraint is relatively easy to enforce for the static-analysis features considered in our study, it can be far more challenging when dealing with dynamic features~\cite{kapratwar2017static,liu2020review}.

\textbf{Duplicate detection.} We treat duplicates as samples that are indistinguishable in the model’s feature space. This does not guarantee duplication in the raw input, since obfuscation, packing, and feature drift can map distinct binaries to identical features~\cite{o2011obfuscation,yang2015appspear,chen2023overkill}. Accurate malware deduplication is difficult and outside our scope; our aim is to assess reproducibility under fixed feature sets and models. Because duplicate handling is induced by the feature extractor rather than the learner, fair model comparison requires deduplication on a common feature set or, at a minimum, explicit reporting of the resulting performance variance.


\textbf{Label Noise and Obfuscation.}  
Datasets with labels aggregated from sources such as VirusTotal may suffer from noise and inconsistencies in ground truth annotations, as noted in prior studies~\cite{aghakhani2020malware}. Similarly, malware developers often employ obfuscation techniques to evade detection. Both issues can affect model performance and bias comparisons across methods. Addressing these challenges is beyond the scope of our work; however, since our focus is on the reproducibility of machine learning models, we believe our findings remain applicable in such settings.





\section{Conclusion}


In this paper, we revisited key reproducibility and replicability challenges in static feature-based Android malware detection. We identified pitfalls in dataset curation, model selection, and evaluation, showing that issues such as data duplication, limited hyperparameter tuning, and biased baselines can cause major performance disparities. Through extensive experiments across models in offline and continuous active learning, we found that well-tuned simple models often match or surpass complex architectures. To promote transparency and fairness, we release an open-source framework for standardized, reproducible benchmarking of new models.

\clearpage

\bibliographystyle{IEEEtran}
\bibliography{main}

\clearpage

\appendix

\subsection{Implementation Details}

\subsubsection{Environment}
We conducted all experiments using Python 3.11.6 on Red Hat Enterprise Linux (Version 9.4). Neural network training was performed using a single NVIDIA A100 GPU. Table~\ref{tab-pip} lists the Python pip packages used in the implementation:

\begin{table}[h]
\centering
\small
\begin{tabular}{llp{4cm}} 
\toprule
\textbf{Pip Package} & \textbf{Version} & \textbf{Utility} \\ \midrule
scikit-learn & 1.5.0 & Used to train RF, SVM \\ 
torch & 2.1.0 & Used to train neural networks (MLP, SCC, HCC) \\ 
xgboost & 2.1.0 & Used to train XGBoost \\ 
numpy & 1.26.4 & Used to process datasets \\ 
pytorch-metric-learning & 2.6.0 & Used for creating data sampler for HCC \\ 
yacs & 0.1.8 & Configuration management for experiments \\ 
\bottomrule
\end{tabular}
\caption{Python pip packages used in the implementation}
\label{tab-pip}
\end{table}

\subsubsection{Code Structure}
The code is organized into multiple key modules to facilitate easy extension:

\begin{enumerate}
    \item \textbf{Dataset:} We preprocess data into a single \texttt{.npz} file containing features, labels, timestamps, and duplication indicators. The \texttt{intra\_split\_dupes} field marks duplicate within the current timestamp, referring to earlier indices or set to \texttt{-1} if unique. The \texttt{cross\_split\_dupes} field tracks duplicates across splits, indicating the relevant split and index or \texttt{none} if unique. This format supports active and offline learning deduplication. The \texttt{MalwareDataset} class handles dataset loading and includes the HalfSampler method used in HCC~\cite{chen2023continuous}, as well as a class for working with Triplet Datasets.

    \item \textbf{Models:} This module implements six models, all inheriting from a \texttt{BaseModel} class, which takes \texttt{params} for hyperparameters and \texttt{cfg} for experiment-specific settings (e.g., output classes, GPU usage). Each model includes methods for fitting (\texttt{fit}), prediction (\texttt{predict}, \texttt{predict\_proba}), and active learning sample selection (\texttt{sample\_active\_learning}).

    \item \textbf{Tasks:} We separate offline and active learning into distinct tasks within the \texttt{tasks} module, each with specific training and evaluation procedures. Tasks use a \texttt{yacs} configuration~\cite{yacs}, provided as a YAML file, specifying the task type, model, and relevant settings. This separation allows easy extension for other malware analysis tasks, such as family classification or concept drift detection, with minimal code changes.
\end{enumerate}

\section{Hyperparameter Search} \label{appendix-hpo}

\subsection{Hyperparameter Search-Space}

The detailed hyperparameter search space is shown in Table~\ref{tab:hyperparameters}. Most hyperparameter names follow standard terminologies in the scikit-learn, XGBoost, or PyTorch APIs. Below, we describe those with different or new terminology:

\begin{itemize}
    \item \textbf{XGBoost/MLP/SCC - class-weight:} This parameter, "scale\_pos\_weight" in XGBoost, defaults to 1 if set to False. When True, it adjusts for class imbalance by weighting the positive class (malware) according to the benign-to-malware ratio in the training data. For MLP and SCC, samples are either balanced in each batch with equal benign and malware samples (balance=True) or drawn randomly (balance=False).

    \item \textbf{MLP/SCC/HCC - cont\_learning\_epochs:} In continuous active learning with neural networks, initial and retraining epochs differ, similar to~\cite{chen2023continuous}. For HCC, it specifies the number of epochs; for MLP and SCC, it represents the fraction of the original training epochs.

    \item \textbf{HCC - cont\_learning\_lr:} For HCC, the initial learning rate is varies the learning rate for each update, while MLP and SCC continue with the current rate using the Adam optimizer.

    \item \textbf{SCC/HCC - xent\_lambda:} This controls the binary cross-entropy loss weight in contrastive learning. We use a value of 100, as suggested in~\cite{chen2023continuous}.
    
    \item \textbf{SCC/HCC - margin:} This sets the margin for computing contrastive loss. We use a value of 10, following~\cite{chen2023continuous}.
\end{itemize}

We adopt strategies from prior works on Android malware detection~\cite{chen2023continuous}, machine learning on tabular data~\cite{mcelfresh2024neural}, and deep neural architectures~\cite{bengio2012practical} to design the hyperparameter search space.

While we mostly follow the hyperparameters for HCC from~\cite{chen2023continuous}, there are a few minor differences. These choices ensure consistency with other neural network architectures and ease of implementation. We explore two architectures for the encoder and MLP layers, each instead of the single one used in their work. We include dropout as a hyperparameter instead of setting it to a fixed value of 0.2 in their study. Only the Adam optimizer is used during the continuous learning retraining phase. We employ a step learning rate scheduler and omit the cosine annealing scheduler. Regardless, we verify that our search space includes the best hyperparameter set reported in~\cite{chen2023continuous}; for instance, cosine annealing did not outperform step-based learning rates in any dataset in their results.

\begin{table*}[h]
\caption{Hyperparameter Search Spaces for Different Models. U indicates sampling from a uniform random distribution over values in the range. * indicates hyperparameters that are only used during continuous learning retraining phase.}
\centering
\resizebox{0.95\textwidth}{!}{%
\begin{tabular}{p{3cm} p{7cm} p{7cm}} 
\toprule
\textbf{Model} & \textbf{Hyperparameter} & \textbf{Candidate Values} \\
\midrule
\multirow{4}{*}{Random Forest} & n\_estimators &  $2^x$, where $x \in \text{U}[5, 10]$ \\
 & max\_depth & $2^y$, where $y \in \text{U}[5, 10]$ \\
 & criterion &  \{gini, entropy, log\_loss\} \\
 & class\_weight & \{None, "balanced"\} \\
\midrule
\multirow{2}{*}{SVM} & C & $10^z$, where $z \in \text{U}[-4, 3]$ \\
 & class\_weight & \{None, "balanced"\} \\
\midrule
\multirow{6}{*}{XGBoost} & max\_depth & $2^w$, where $w \in \text{U}[3, 7]$ \\
 & alpha & $10^a$, where $a \in \text{U}[-8, 0]$ \\
 & lambda & $10^b$, where $b \in \text{U}[-8, 0]$ \\
 & eta & $3.0 \times 10^c$, where $c \in \text{U}[-2, -1]$ \\
 & balance & \{True, False\} \\
 & num\_boost\_round & \{100, 150, 200, 300, 400\} \\ 
& subsample & $x$, where $x \in \text{U}[0.8, 1.0]$ \\
& colsample\_bytree & $x$, where $x \in \text{U}[0.8, 1.0]$ \\

\midrule
\multirow{7}{*}{MLP} & mlp\_layers & \{[100, 100], [512, 256, 128], [512, 384, 256, 128], [512, 384, 256, 128, 64]\} \\
 & learning\_rate & $10^d$, where $d \in \text{U}[-5, -3]$ \\
 & dropout & $x$, where $x \in \text{U}[0.0, 0.5]$ \\
 & batch\_size & $2^e$, where $e \in $ \{5, 6, 7, 8, 9, 10\} \\
 & epochs & \{25, 30, 35, 40, 50, 60, 80, 100, 150\} \\
 & optimizer & \{Adam\} \\
 & balance & \{True, False\} \\
 & cont\_learning\_epochs* & \{0.1, 0.2, 0.3, 0.4, 0.5\} \\
\midrule
\multirow{10}{*}{SCC} & encoder\_layers & \{[512, 256, 128], [512, 384, 256, 128]\} \\
 & mlp\_layers & \{[100], [100, 100]\} \\
 & learning\_rate & $10^f$, where $f \in \text{U}[-5, -3]$ \\
 & dropout & $y$, where $y \in \text{U}[0.0, 0.25]$ \\
 & batch\_size & $2^g$, where $g \in $ \{9, 10, 11\} \\
 & epochs & \{25, 30, 35, 40, 50, 60, 80, 100\} \\
 & xent\_lambda & \{100\} \\
 & margin & \{10\} \\
 & optimizer & \{Adam\} \\
 & balance & \{True, False\} \\
 & cont\_learning\_epochs* & \{0.1, 0.2, 0.3, 0.4, 0.5\} \\
\midrule
\multirow{12}{*}{HCC} & encoder\_layers & \{[512, 256, 128], [512, 384, 256, 128]\} \\
 & mlp\_layers & \{[100], [100, 100]\} \\
 & learning\_rate & \{0.001, 0.003, 0.005, 0.007\} \\
 & dropout & $z$, where $z \in \text{U}[0.0, 0.25]$ \\
 & batch\_size & $2^{10}$ \\
 & epochs & \{100, 150, 200, 250\} \\
 & xent\_lambda & \{100\} \\
 & margin & \{10\} \\
 & optimizer & \{Adam, SGD\} \\
 & scheduler\_step & \{10\} \\
 & scheduler\_gamma & \{0.5, 0.95\} \\
 & cont\_learning\_lr* & \{0.01, 0.05\} \\
 & cont\_learning\_epochs* & \{50, 100\} \\
\bottomrule
\end{tabular}
}

\label{tab:hyperparameters}
\end{table*}


\begin{figure*}
\centering
\includegraphics[width=0.45\textwidth]{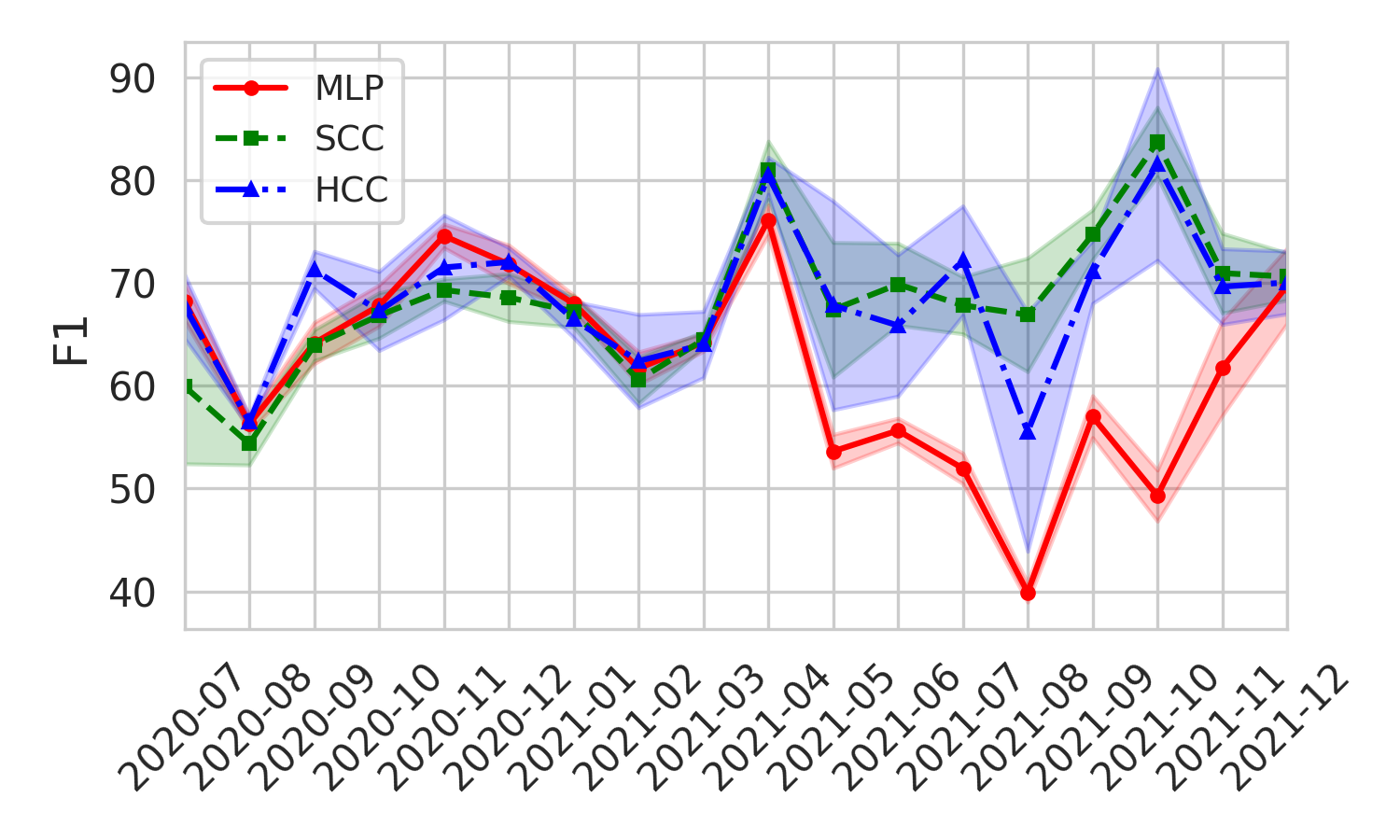}
\includegraphics[width=0.45\textwidth]{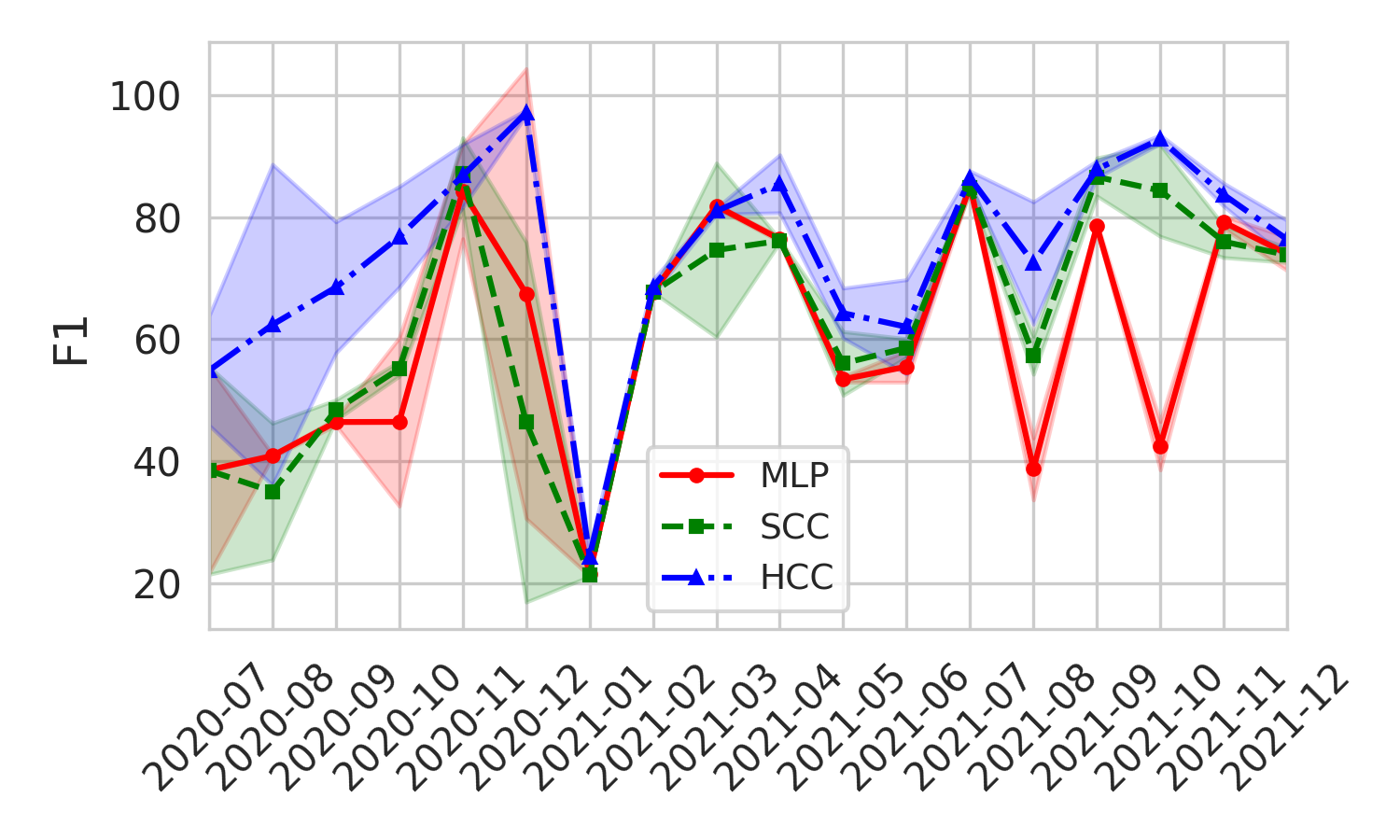}

\caption{F1-score over the test months for active learning on the (left) deduplicated (right) duplicated Drebin datasets}
\label{fig:al-variance}
\end{figure*}

\subsection{Variance in Results for Continuous Learning}

Figure~\ref{fig:al-variance} shows the monthly F1 scores on the Drebin dataset in continuous active learning settings, comparing three neural networks on both duplicated and deduplicated datasets. Separate hyperparameter tuning was performed for MLP and SCC, while for HCC, we used the artifacts provided in~\cite{chen2023continuous} for the duplicated setting to maintain consistency with their study. Notably, we observe extreme variance in model performance in certain months (e.g., December 2020) in the duplicated datasets.

\subsection{Feature Index Mapping}
\label{app:mapping}
Tables~\ref{feat-api} and~\ref{feat-dre} show the feature index mappings for the important features used by different models on the APIGraph and Drebin datasets, respectively.

\begin{table}[h]
\centering
\caption{APIGraph API feature mapping}
\label{tab:suspicious_features}
\begin{tabular}{|c|>{\raggedright\arraybackslash}p{0.72\linewidth}|}
\hline
Feature ID & Feature Name \\ \hline
2 & \path{usedpermissionslist_android.permission.get_tasks} \\ \hline
4 & \path{usedpermissionslist_android.permission.read_phone_state} \\ \hline
12 & \path{suspiciousapilist_landroid/telephony/telephonymanager.getdeviceid} \\ \hline
14 & \path{suspiciousapilist_system/bin/su} \\ \hline
16 & \path{suspiciousapilist_landroid/telephony/telephonymanager.getsubscriberid} \\ \hline
17 & \path{suspiciousapilist_landroid/content/pm/packagemanager.getpackageinfo} \\ \hline
21 & \path{urldomainlist_10.0.0.172} \\ \hline
31 & \path{activitylist_com.waps.offerswebview} \\ \hline
37 & \path{requestedpermissionlist_android.permission.receive_boot_completed} \\ \hline
41 & \path{requestedpermissionlist_android.permission.get_tasks} \\ \hline
44 & \path{requestedpermissionlist_android.permission.read_phone_state} \\ \hline
46 & \path{suspiciousapilist_landroid/telephony/gsm/smsmanager.sendtextmessage} \\ \hline
68 & \path{usedpermissionslist_android.permission.send_sms} \\ \hline
75 & \path{suspiciousapilist_landroid/telephony/smsmanager.sendtextmessage} \\ \hline
87 & \path{requestedpermissionlist_android.permission.send_sms} \\ \hline
102 & \path{intentfilterlist_android.intent.action.package_added} \\ \hline
107 & \path{intentfilterlist_android.intent.action.package_removed} \\ \hline
131 & \path{requestedpermissionlist_com.android.launcher.permission.install_shortcut} \\ \hline
133 & \path{requestedpermissionlist_com.android.browser.permission.read_history_bookmarks} \\ \hline
150 & \path{intentfilterlist_android.intent.action.user_present} \\ \hline
154 & \path{requestedpermissionlist_android.permission.read_contacts} \\ \hline
157 & \path{usedpermissionslist_android.permission.get_accounts} \\ \hline
192 & \path{requestedpermissionlist_android.permission.system_alert_window} \\ \hline
197 & \path{requestedpermissionlist_android.permission.call_phone} \\ \hline
\end{tabular}
\label{feat-api}
\end{table}

\begin{table}[h]
\centering
\tiny
\caption{Drebin API feature mapping}
\label{tab:suspicious_features}
\begin{tabular}{|c|>{\raggedright\arraybackslash}p{0.72\linewidth}|}
\hline
Feature ID & Feature Name \\ \hline
1 & \path{suspiciousapilist_android.content.res.assetmanager.open} \\ \hline
2 & \path{suspiciousapilist_android.content.context.getassets} \\ \hline
4 & \path{suspiciousapilist_android.app.application.oncreate} \\ \hline
6 & \path{intentfilterlist_com.android.vending.install_referrer} \\ \hline
7 & \path{intentfilterlist_android.intent.action.boot_completed} \\ \hline
8 & \path{requestedpermissionlist_android.permission.read_external_storage} \\ \hline
9 & \path{requestedpermissionlist_android.permission.receive_boot_completed} \\ \hline
10 & \path{requestedpermissionlist_android.permission.system_alert_window} \\ \hline
11 & \path{requestedpermissionlist_android.permission.get_tasks} \\ \hline
12 & \path{requestedpermissionlist_android.permission.write_settings} \\ \hline
13 & \path{requestedpermissionlist_android.permission.access_wifi_state} \\ \hline
14 & \path{requestedpermissionlist_android.permission.read_phone_state} \\ \hline
15 & \path{requestedpermissionlist_android.permission.change_wifi_state} \\ \hline
16 & \path{requestedpermissionlist_android.permission.write_external_storage} \\ \hline
17 & \path{suspiciousapilist_android.app.alertdialog\$builder.setcancelable} \\ \hline
18 & \path{suspiciousapilist_android.app.alertdialog\$builder.setmessage} \\ \hline
21 & \path{suspiciousapilist_android.view.window.settype} \\ \hline
22 & \path{suspiciousapilist_android.content.res.resources.getassets} \\ \hline
25 & \path{intentfilterlist_android.intent.action.view} \\ \hline
26 & \path{requestedpermissionlist_android.permission.access_coarse_location} \\ \hline
27 & \path{requestedpermissionlist_android.permission.camera} \\ \hline
28 & \path{requestedpermissionlist_android.permission.access_fine_location} \\ \hline
29 & \path{requestedpermissionlist_android.permission.vibrate} \\ \hline
35 & \path{usedpermissionslist_android.permission.access_network_state} \\ \hline
44 & \path{suspiciousapilist_android.app.application.getpackagename} \\ \hline
182 & \path{suspiciousapilist_android.content.context.getsharedpreferences} \\ \hline
904 & \path{suspiciousapilist_android.app.application.registeractivitylifecyclecallbacks} \\ \hline
1468 & \path{suspiciousapilist_android.content.intent.setflags} \\ \hline
1615 & \path{suspiciousapilist_android.widget.toast.show} \\ \hline
2345 & \path{suspiciousapilist_android.content.pm.packagemanager\$namenotfoundexception.printstacktrace} \\ \hline
2744 & \path{suspiciousapilist_android.telephony.telephonymanager.getdeviceid} \\ \hline
3279 & \path{suspiciousapilist_ljava/io/ioexception;->printstacktrace} \\ \hline
3727 & \path{suspiciousapilist_android.net.wifi.wifiinfo.getmacaddress} \\ \hline
4256 & \path{suspiciousapilist_android.content.context.getdir} \\ \hline
4665 & \path{suspiciousapilist_android.view.layoutinflater.from} \\ \hline
5358 & \path{suspiciousapilist_ljava/lang/runtime;->exec} \\ \hline
6216 & \path{intentfilterlist_com.google.android.c2dm.intent.receive} \\ \hline
6218 & \path{contentproviderlist_android.support.v4.content.fileprovider} \\ \hline
6219 & \path{hardwarecomponentslist_android.hardware.camera.autofocus} \\ \hline
6220 & \path{hardwarecomponentslist_android.hardware.camera} \\ \hline
6274 & \path{suspiciousapilist_android.view.surfaceview.<init>} \\ \hline
6402 & \path{suspiciousapilist_android.app.activitymanager.getrunningservices} \\ \hline
6405 & \path{suspiciousapilist_android.content.pm.packagemanager.getinstalledpackages} \\ \hline
6475 & \path{suspiciousapilist_android.content.pm.applicationinfo.loadlabel} \\ \hline
6589 & \path{activitylist_com.google.android.gms.ads.adactivity} \\ \hline
6591 & \path{activitylist_com.google.android.gms.common.api.googleapiactivity} \\ \hline
6592 & \path{requestedpermissionlist_android.permission.get_accounts} \\ \hline
12412 & \path{requestedpermissionlist_com.android.vending.billing} \\ \hline
15566 & \path{suspiciousapilist_android.content.pm.packagemanager.setcomponentenabledsetting} \\ \hline
16973 & \path{contentproviderlist_mono.monoruntimeprovider} \\ \hline
\end{tabular}
\label{feat-dre}
\end{table}

\end{document}